\documentclass[twocolumn]{aastex61}
\usepackage[utf8]{inputenc}
\usepackage{natbib} 
\usepackage{xspace}    
\usepackage{amsmath}
\usepackage{color}

\newcommand{\modif}[1]{#1}
\newcommand{\firstD}{HD~104860\xspace}
\newcommand{\secD}{HD~192758\xspace}

\received{2017 December 16}
\revised{2018 January 13}
\accepted{2018 January 15}
\submitjournal{ApJ}
\shorttitle{Debris disks imaged around HD~104860 and HD~192758}
\shortauthors{Choquet et al.}

\begin{document}

\title{HD~104860 and HD~192758: two debris disks newly imaged in scattered-light with HST}

\correspondingauthor{\'E.~Choquet}
\email{echoquet@jpl.nasa.gov}%

\author[0000-0002-9173-0740]{\'E.~Choquet}\altaffiliation{Hubble Fellow}
\affiliation{Department  of  Astronomy,  California  Institute  of  Technology, 1200 E. California Blvd, Pasadena, CA 91125, USA}
\affiliation{Jet Propulsion Laboratory, California Institute of Technology, 4800 Oak Grove Drive, Pasadena, CA 91109, USA}
\author{G.~Bryden}
\affiliation{Jet Propulsion Laboratory, California Institute of Technology, 4800 Oak Grove Drive, Pasadena, CA 91109, USA}
\author{M.~D.~Perrin}
\affiliation{Space Telescope Science Institute, 3700 San Martin Dr, Baltimore MD 21218, USA}
\author{R.~Soummer}
\affiliation{Space Telescope Science Institute, 3700 San Martin Dr, Baltimore MD 21218, USA}
\author{J.-C.~Augereau} 
\affiliation{Univ. Grenoble Alpes, CNRS, IPAG, 38000 Grenoble, France}
\author{C.~H.~Chen}
\affiliation{Space Telescope Science Institute, 3700 San Martin Dr, Baltimore MD 21218, USA}
\author{J.~H.~Debes}
\affiliation{Space Telescope Science Institute, 3700 San Martin Dr, Baltimore MD 21218, USA}
\author{E.~Gofas-Salas}
\affiliation{ONERA, The French Aerospace Lab, 92322 Ch\^atillon, France}
\affiliation{Institut de la Vision, 75012 Paris, France}
\author{J.~B.~Hagan}
\affiliation{Space Telescope Science Institute, 3700 San Martin Dr, Baltimore MD 21218, USA}
\author{D.~C.~Hines}
\affiliation{Space Telescope Science Institute, 3700 San Martin Dr, Baltimore MD 21218, USA}
\author{D.~Mawet}
\affiliation{Department  of  Astronomy,  California  Institute  of  Technology, 1200 E. California Blvd, Pasadena, CA 91125, USA}
\affiliation{Jet Propulsion Laboratory, California Institute of Technology, 4800 Oak Grove Drive, Pasadena, CA 91109, USA}
\author{F.~Morales}
\affiliation{Jet Propulsion Laboratory, California Institute of Technology, 4800 Oak Grove Drive, Pasadena, CA 91109, USA}
\author{L.~Pueyo}
\affiliation{Space Telescope Science Institute, 3700 San Martin Dr, Baltimore MD 21218, USA}
\author{A.~Rajan}
\affiliation{Space Telescope Science Institute, 3700 San Martin Dr, Baltimore MD 21218, USA}
\author{B.~Ren}
\affiliation{Johns Hopkins University, 3400 North Charles Street, Baltimore, MD 21218, USA}
\author{G.~Schneider}
\affiliation{Steward Observatory, The University of Arizona, 933 North Cherry Avenue, Tucson, AZ 85721, USA}
\author{C.~C.~Stark}
\affiliation{Space Telescope Science Institute, 3700 San Martin Dr, Baltimore MD 21218, USA}
\author{S.~Wolff}
\affiliation{Leiden Observatory, Leiden University, 2300 RA Leiden, The Netherlands}

\begin{abstract}
We present the first scattered-light images of two debris disks around the F8 star \firstD and the F0V star \secD, respectively $\sim45$ and $\sim67$~pc away. We detected these systems in the F110W and F160W filters through our re-analysis of archival \emph{Hubble Space Telescope} NICMOS data with modern starlight subtraction techniques. Our image of \firstD confirms the morphology previously observed by Herschel in thermal emission with a well-defined ring at radius $\sim114$~au inclined $\sim58$\degr{}. Although the outer edge profile is consistent with dynamical evolution models, the sharp inner edge suggests sculpting by unseen perturbers. Our images of \secD reveal a disk at radius $\sim95$~au inclined by $\sim59$\degr{}, never resolved so far. These disks have low scattering albedos of 10\% and 13\% respectively, inconsistent with water ice grain compositions. They are reminiscent of several other disks with similar inclination and scattering albedos: Fomalhaut, HD~92945, HD~202628, and HD~207129. They are also very distinct from brighter disks in the same inclination bin, which point to different compositions between these two populations. Varying scattering albedo values can be explained by different grain porosities, chemical compositions, or grain size distributions, which may indicate distinct formation mechanisms or dynamical processes at work in these systems. 
Finally, these faint disks with large infrared excesses may be representative of an underlying population of systems with  low albedo values. Searches with more sensitive instruments on HST or on the \emph{James Webb Space Telescope} and using state-of-the art starlight-subtraction methods may help discover more of such faint systems.

\end{abstract}


\section{Introduction}

Debris disks are extrasolar system components evolving around main-sequence stars. They are composed of kilometer-sized planetesimals formed during the earlier protoplanetary stage of the system, and of dust particles generated by colliding bodies through a destructive grinding cascade stirred by secular perturbations from planets or large planetesimals \citep[see][for a review]{Wyatt2008}. 
About five hundreds debris disk systems have been identified around nearby stars from their photometric excess in the infrared \citep{Eiroa2013,Chen2014, Patel2017}, revealing that massive dust systems are as common as 20 to 26\% around A to K type stars in the solar neighborhood \citep{Thureau2014,Montesinos2016}.  

Debris disks are the perfect laboratories to study the dynamical balance ruling circumstellar environments. Particles of different sizes, masses, and distances from their host stars are affected differently by the stellar radiation, wind, gravity, and by perturbing bodies like sub-stellar companions \citep{Krivov2010}. Gravity is the dominant force for planetesimals and millimeter-sized grains, which orbit close to their parent bodies. Small particles, having a large cross-section compared to their volume, are on the other hand very sensitive to the radiative pressure and drag forces, and spread both inward and outward from their parent bodies on eccentric orbits. The smallest particles (sub-micron size) are blown out of the system by the stellar radiative pressure, resulting in an abrupt cut-off to the size distribution in debris disks and to their slow mass-decay  with time.

Resolved images of debris disks in several wavelength regimes provide  ideal probes to study dynamics in circumstellar environments. Images reveal the disks morphology and the spatial distribution of their dust, and multi-wavelength imagery additionally traces particles of different sizes. As millimeter-sized grains are the most efficient emitters at long wavelengths, far-IR/millimeter images map the spatial distribution of large grains thermal emission, and, indirectly, the location of their planetesimals parent bodies \citep[e.g.][]{Booth2016,Booth2017}. Micron-sized particles are inefficient emitters, but efficiently scatter the starlight at wavelengths comparable to their size. Near-infrared and visible-light imaging thus shows how small dust grains spread out in debris disk systems \citep[e.g.][]{Schneider2014}. By characterizing the dust spatial and size distribution, multi-spectral imaging enables us to study the dynamical balance in disks as a function on the stellar environment \citep[e.g.][]{MacGregor2017}. Furthermore, resolved images let us study possible planet disk interactions by characterizing the imprints of unseen planets on disks morphologies \citep{Lee2016}. Giant gaseous planets, although rare at 10--100~au, are more commonly found in systems harboring massive debris disks \citep{Meshkat2017}.

\begin{figure}
\center
\includegraphics[width=8.5cm]{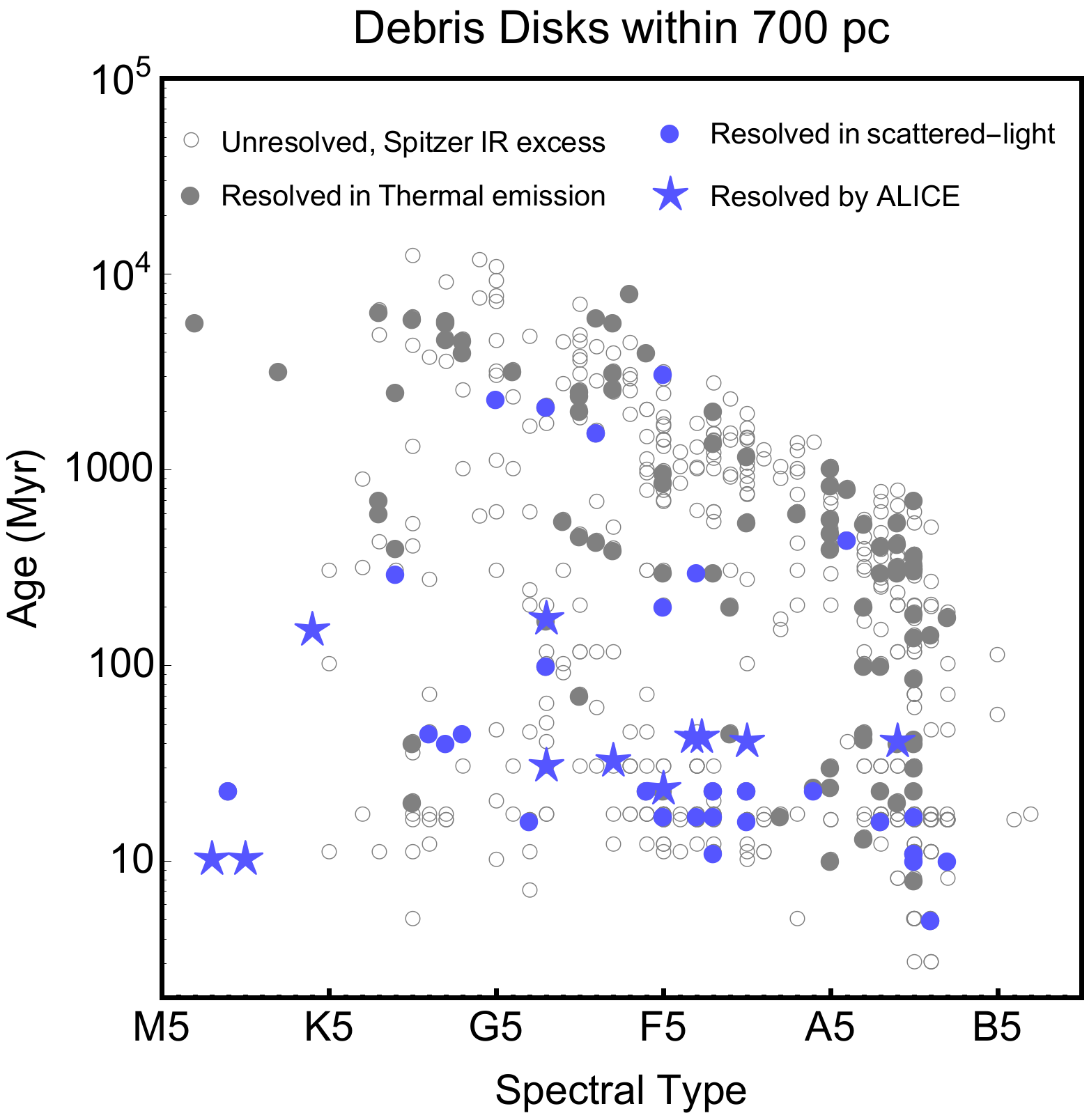}
\caption{Age and spectral type distribution of the 500 stars within 700~pc with a debris disk detected in thermal emission by the Spitzer Space telescope \citep[grey empty circles][]{Chen2014}, and/or that have been resolved in thermal emission (grey filled circles, 97 systems) or in scattered-light (blue markers, 41 systems). The blue stars highlight the debris disks firstly imaged by ALICE in scattered-light. We note that the Spitzer sample is composed at 94\% of stars within 250~pc, with only a few systems beyond 500~pc. \label{fig:alldisks}}
\end{figure}

About a hundred debris disks have been spatially resolved to date, mostly in thermal emission at 70--160~$\mu$m with the \emph{Herschel Space Observatory} \citep[e.g.][]{Booth2013,Eiroa2013,Matthews2014a,Morales2016,Vican2016} or at submillimeter/millimeter wavelengths with JCMT  \citep{Holland2017} and ALMA \citep[e.g.][]{MacGregor2013,MacGregor2016a,Su2017}. Using the measured radii of a sample of \emph{Herschel}-resolved disks, \citet{Pawellek2014} interestingly showed that the typical grain size in these disks does not directly scale with the radiative pressure blowout particle size, but decreases with stellar luminosity. This may indicate that other mechanisms are at work in debris disks that limit the production of small grains around late stars, or that induce a higher level of excitation and planetesimal stirring around early stars \citep{Pawellek2015}. 

Such hypotheses could be tested by studying the small dust population in debris disks with scattered-light imaging. However, about 40 debris disks have been resolved so far in scattered-light, and only half of these have also been resolved in thermal emission (see Fig.~\ref{fig:alldisks}). Resolving disks in this wavelength-regime is technically challenging, as the star is typically 1000 times brighter than the total disk emission at these wavelengths, and the disks, typically within a few arcseconds of their host star, are buried within the bright and temporally varying diffraction pattern.

In this paper we report the detection of two new debris disks in the scattered-light regime. They were identified through our re-analysis of archival coronagraphic data from the \emph{Hubble Space Telescope} (HST) NICMOS instrument as part of the \emph{Archival Legacy Investigation of Circumstellar Environments} (ALICE) project\footnote{https://archive.stsci.edu/prepds/alice/}, using modern starlight-subtraction techniques to reveal the faint disk emissions. The disk around \firstD has been previously resolved in thermal emission with Herschel-PACS at 100 and 160~$\mu$m \citep{Morales2013,Morales2016}, and the other, around \secD, has not been resolved so far. These two new detections bring the number of debris disks firstly imaged in scattered-light by the ALICE program to 11, demonstrating the efficacy  of our post-processing method to detect faint resolved circumstellar material \citep{Soummer2014,Choquet2016,Choquet2017}. Along with the 5 debris disk detected during its operational time, NICMOS currently  holds the record of the most debris disks firstly imaged in scattered-light.

In Sec.~\ref{sec:data}, we present the datasets used for this work and the data reduction process. Section~\ref{sec:disks} presents the two systems \firstD and \secD and describes the debris disk detections. We present our analysis of the disk morphologies in Sec.~\ref{sec:analysis}.

\section{Datasets and Data Processing}\label{sec:data}

\subsection{Datasets}

The data on \firstD and \secD  were obtained as part of two surveys with the near-IR NICMOS instrument on HST that aimed at resolving a selection of debris disks identified from their infrared excess, respectively with the \emph{Spitzer} Space Telescope (HST-GO-10527, PI: D. Hines) and with IRAS/Hipparcos (HST-GO-11157, PI: J. Rhee). The data were all obtained with the intermediate-sampling camera NIC2 (0\farcs07565~pixel$^{-1}$) and with its coronagraphic mode featuring a 0\farcs3-radius occulting mask. 

\firstD was observed in two telescope orientations, with the spacecraft rolled by 30\degr{} to enable subtraction of the coronagraphic Point Spread Function (PSF)  through roll differential imaging \citep{Lowrance1999}. \secD was also observed in two 30\degr-different rolls on 2007-07-03, then again in two other rolls about a year later (UT-2008-06-04). However, the guide star acquisition failed during one roll of the later observing sequence, and none of the integrations within this roll were acquired with the star centered on the occulting spot. We only used exposures from the three successful roll acquisitions in our study,  and we combined data from both epochs to maximize the signal to noise ratio (S/N) on the detection, as we do not expect significant temporal variations in the disk morphology and photometry.

\firstD was observed with the NICMOS F110W filter, and \secD with both the F110W and F160W filters. These two wide-band filters were the most commonly used for NICMOS coronagraphic observations, which facilitated assembling large and homogeneous PSF libraries for advanced post-processing \citep{Hagan2018}. The F160W NICMOS filter (pivot wavelength $1.600~\mu$m, 98\%-integrated bandwidth $0.410~\mu$m) is comparable to the ground-based H bandpass, while NICMOS F110W filter (pivot wavelength $1.116~\mu$m, 98\%-integrated bandwidth $0.584~\mu$m) is twice as extended toward shorter wavelengths as the J band.

The  observing parameters and  instrument characteristics of the three datasets (\firstD-F110W, \secD-F110W, and \secD-F160W) are summarized in Table~\ref{tab:data}.

\subsection{Data Processing}\label{sec:processing}

\begin{deluxetable}{lrrr}
\tablecaption{Observing and processing parameters\label{tab:data}}
\tablehead{ 
\colhead{Parameters}	& \colhead{\firstD}	&\colhead{\secD} &\colhead{\secD}\\
\colhead{}	& \colhead{(F110W)}	&\colhead{(F110W)} &\colhead{(F160W)}
}
\startdata
UT date				& 2006-03-20	&\multicolumn{2}{c}{2007-07-03}\\
					& 			&\multicolumn{2}{c}{2008-06-04}\\
\# orientations			&2			&\multicolumn{2}{c}{3}\\
Orient difference (\degr)	&30			&\multicolumn{2}{c}{30; 2; 28}\\
Filter					& F110W		&F110W	&F160W	\\
$\lambda_p$ ($\mu$m)& 1.116		&1.116	&1.600	\\
$F_{\nu}$ ($\mu$Jy.s.DN$^{-1}$)&1.21121&1.21121	&1.49585\\
\#  combined frames		&18			&15		&15		\\
Total exp. time (s)		&5183		&3456	&3456 	\\
\# frames PSF library	&439			&327		&277 	\\
\# subtracted PCs   		&141			&130		&80		\\
Image crop size (pix.)	&140			&80		&80		\\
Mask radius (pix.)		&13		&8		&8\\
\enddata
\end{deluxetable} 

To detect the faint signal of the disks compared to the star, we used the multi-reference star differential imaging (MRDI) PSF-subtraction method developed for the ALICE program \citep{Soummer2011,Choquet2014d}. We assembled and registered large and homogeneous libraries of coronagraphic images from the NICMOS archive, gathering images from multiple reference stars observed as part of several HST programs. These libraries were used to subtract the star PSF from each exposure of the science target with the Principal Component Analysis (PCA) KLIP algorithm \citep{Soummer2012}.  

All the images used to process the \firstD dataset (PSF library and science images) were calibrated with contemporary flat-field images and observed dark frames as part of the \emph{Legacy Archive PSF Library and Circumstellar Environments} (LAPLACE) program\footnote{https://archive.stsci.edu/prepds/laplace}. Conversely, \secD's data were obtained in HST cycle 16, last operational cycle of NICMOS, and were not included in the LAPLACE calibration program. To have libraries large enough to sample the PSF variations as well as representative of \secD's images, we assembled PSF libraries both from LAPLACE programs and from the non-LAPLACE  HST-GO-11157 program. The later was processed with the NICMOS \textit{calnica} calibration pipeline and  bad pixel corrected. 
For \firstD's and \secD's datasets, we respectively down-selected the 60\% and the 30\% of the images in the libraries the most correlated with their respective science frames. This selective criterion ensures that the images in the libraries are well representative of the star PSF in each science frames \citep{Choquet2014d}, and minimizes over-subtraction of the astrophysical signal, especially in the case of extended objects which project more strongly on poorly-matched PSF components with PCA-type algorithms. 

We applied the KLIP algorithm on the cropped sub-images of the science target excluding a central circular area. The numbers of principal components (PCs) used for the PSF subtraction were selected to maximize the S/N on the disks after visual inspection of the final images. The PSF-subtracted exposures were then rotated to have North pointing up, co-added, and scaled  to surface brightness units based on the exposure time, HST's calibrated photometric factors ($F_{\nu}$), and plate scale. The post-processing parameters (crop size, mask radius, PSF library size, number of PCs) are listed in Table~\ref{tab:data}.

To estimate the noise in  the final images, we processed the reference star images from the PSF libraries with the same method and reduction parameters as the science images. The PSF-subtracted libraries were then partitioned into sets with the same number of frames as the science targets, rotated with the target image orientations, and combined. The noise maps were computed from the pixel-wise standard deviation across these sets of processed reference star images. 

\begin{deluxetable}{lrr}
\tablecaption{System properties  \label{tab:stars}}
\tablehead{
 \colhead{Properties}	& \colhead{\firstD}	&\colhead{\secD}
 }
\startdata
RA (J2000) 	&  12 04 33.731	&  20 18 15.790	\\
DEC (J2000)	&+66 20 11.715	&-42 51 36.297		\\
Spectral Type	&F8				&F0V		\\
J (mag)		&6.822 (1)			&6.387 (1)	\\
H (mag)		&6.580  (1)		&6.298 (1)	\\
Distance (pc)	&$45.0\pm0.5$ (2)	&$67\pm2$ (2)	\\
PM RA (\arcsec/yr)	&$-57.07\pm0.06$ (2)	&$51.65\pm0.07$ (2)\\
PM DEC (\arcsec/yr)	&$43.7\pm0.3$ (2)		&$-57.9\pm0.4$ (2)\\
Age (Myr)		&19--635	(3,4)		&45--830 (5,6)	\\
Association	&Field (4)			&Field / ICS 2391 (5,6)\\
$L_{dust}/L_\star$&$6.3\times10^{-4}$ (3)	&$5.7\times10^{-4}$	(7)\\
\enddata
\tablerefs{1: \citealt{Cutri2003}; 2: \citealt{GaiaCollaboration2016}; 3: \citealt{Hillenbrand2008}; 4: \citealt{Brandt2014b}; 5: \citealt{Moor2006}; 6: \citealt{Chen2014}; 7: \citealt{Moor2011a}.}
\end{deluxetable}

\section{Disk detections}\label{sec:disks}
\begin{figure*}[ht]
\center
\includegraphics[width=18cm]{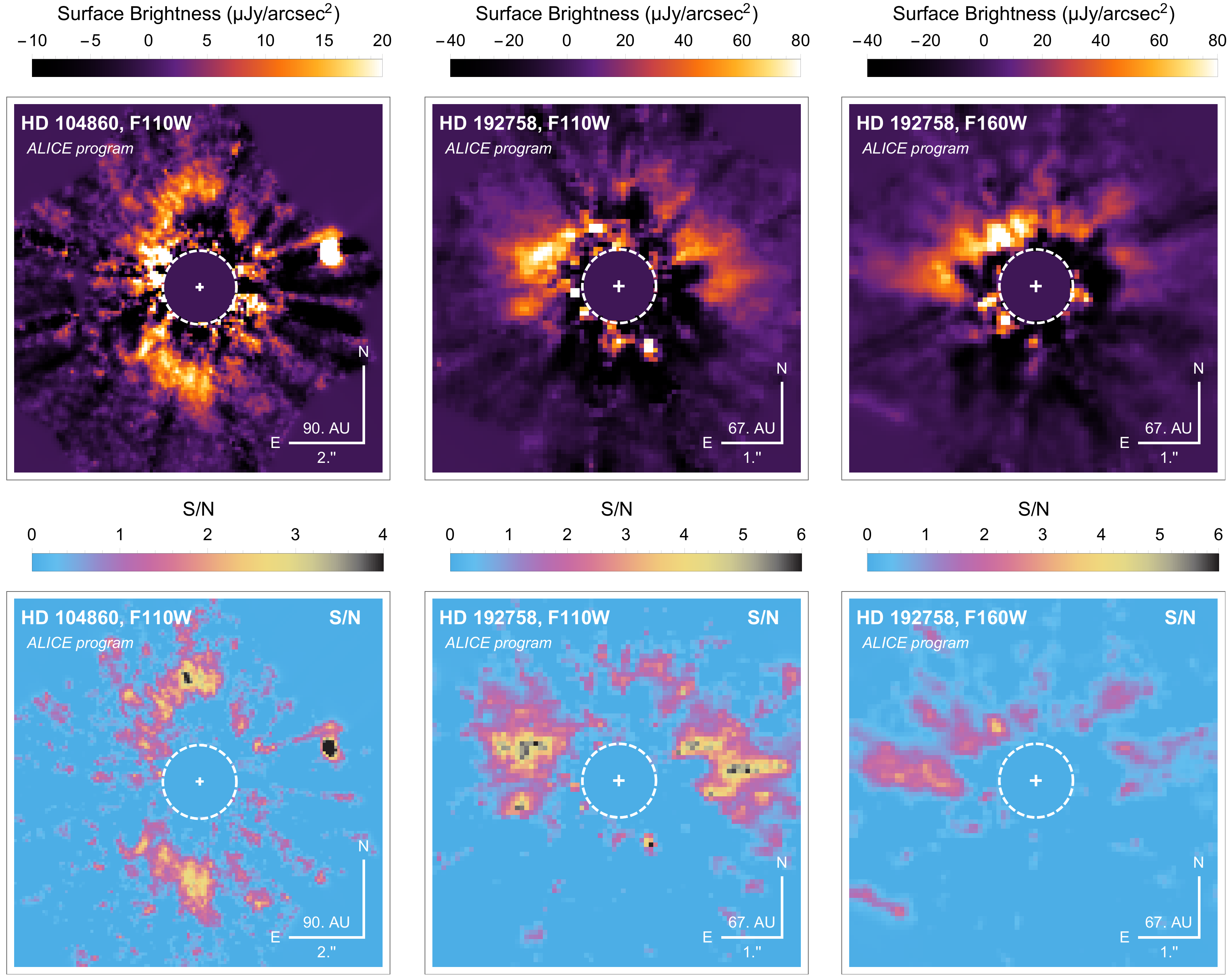}
\caption{Debris disks detected around \firstD (F110W filter) and \secD (F110W and F160W filters) by re-analyzing archival HST-NICMOS data as part of the ALICE program, using libraries built of multiple reference star images from the NICMOS archive and using the PCA-KLIP algorithm. The top row shows the disks combined images in surface brightness units, and the bottom row shows the S/N maps. All images have been smoothed by convolution with a synthetic PSF. The white dashed circles show the area masked for post-processing with KLIP.\label{fig:images}}
\end{figure*}

We detect  faint and resolved dust emission around both stars. Fig.~\ref{fig:images} presents the images of the two disks and their respective S/N maps.
We list some properties of the two systems in Table~\ref{tab:stars}.

\subsection{\firstD}
\firstD is a F8 field star at $45.0\pm0.5$~pc \citep{GaiaCollaboration2016}. Its age was estimated, based on its chromospheric activity, to 32~Myr by \citet{Hillenbrand2008}, and to 19--635~Myr by \citet{Brandt2014b}. The system has a significant infrared excess at $70~\mu$m identified with \emph{Spitzer} with fractional infrared luminosity of $L_{dust}/L_{\star}\sim6.3\times10^{-4}$ \citep{Hillenbrand2008}. Its SED is well-described by a two-temperature black-body model, with a warm dust population at $210\pm27$~K with a fractional luminosity of $2.4\times10^{-5}$ and a cold dust population at $42\pm5$~K with a large infrared fractional luminosity of $2.8\times10^{-4}$ \citep{Chen2014}. Without prior on the dust separation to the star and assuming that the disks are in radiative and collisional equilibrium, these black-body emissions correspond  respectively to a  $\sim10^{-5}~M_{Moon}$-mass disk at a radius of 3~au from the star, and to a massive $\sim1.5~M_{Moon}$ cold disk at a radius of 366~au, based on silicate spherical grain compositions.
By measuring the cold disk spectral index in the millimeter from VLA and ATCA observations, \citet{MacGregor2016b} estimated a dust size distribution in the system with a power law $q=3.64\pm0.15$, consistent with steady state collisional cascade models \citep{Dohnanyi1969,Pan2012,Gaspar2012}. Assuming a disk composed of astro-silicates \citep{Draine2003}, a stellar luminosity of $L_\star=1.16~L_\sun$ and mass of $M_\star=1.04~M_\sun$, the radiative pressure blowout grain size limit is estimated to $a_{blow}\sim0.4~\mu$m, but the minimum grain size is inferred to $\sim7~\mu$m from joint modeling  of the system's SED and \emph{Herschel} images \citep{Pawellek2014}.

The disk was first resolved in thermal emission at 100 and 160~$\mu$m with \emph{Herschel}-PACS \citep{Morales2013} then marginally resolved at 1.3~mm with CARMA  \citep{Steele2016}. The \emph{Herschel} image at 100~$\mu$m shows a disk of radius $116\pm6$~au, three times smaller than the radius inferred from SED modeling, and inclined by $54\pm7$\degr from face-on with a position angle of $1\pm7$\degr. Such a significant difference in radius between observations and SED-modeling is found for many resolved systems and stresses the need for resolved images to put constraints on debris disks properties. Using the radius inferred from the  \emph{Herschel} image to constrain SED modeling, \citet{Morales2016} found that both pure astro-silicate grains and a mixture of water ice and astro-silicates could compose the dust in the system.

In our F110W NICMOS image, we detect a large ring-shape debris disk around \firstD with a radius of $\sim112$~au ($\sim2.5$\arcsec), inclined by $\sim60$\degr{} from face-on. The disk is detected at S/N $\sim$ 2--4 per pixel and S/N $\sim15$ integrated over the disk area detected above 1.5$\sigma$. This detection enables us to put strong constraints on the disk general morphology (see Sec.~\ref{sec:analysis}), but the S/N is too low to reliably comment on the sub-structures within the disk. 
The observed morphology largely agrees with the Herschel thermal emission image, with a 60 times better angular resolution (95~mas vs. 5.9\arcsec resolution), and provides the first image of the dust in the scattered-light regime. The East side is significantly brighter than the West side, indicative of anisotropic scattering from the dust, and showing the near-side of the disk assuming grains preferentially forward-scattering. 
Our image unambiguously reveals a cleared cavity from the disk's inner edge and down to $\sim45$~au from the star, which we further characterize in Sec.~\ref{sec:analysis}. 

\begin{figure}
\center
\includegraphics[width=8.5cm]{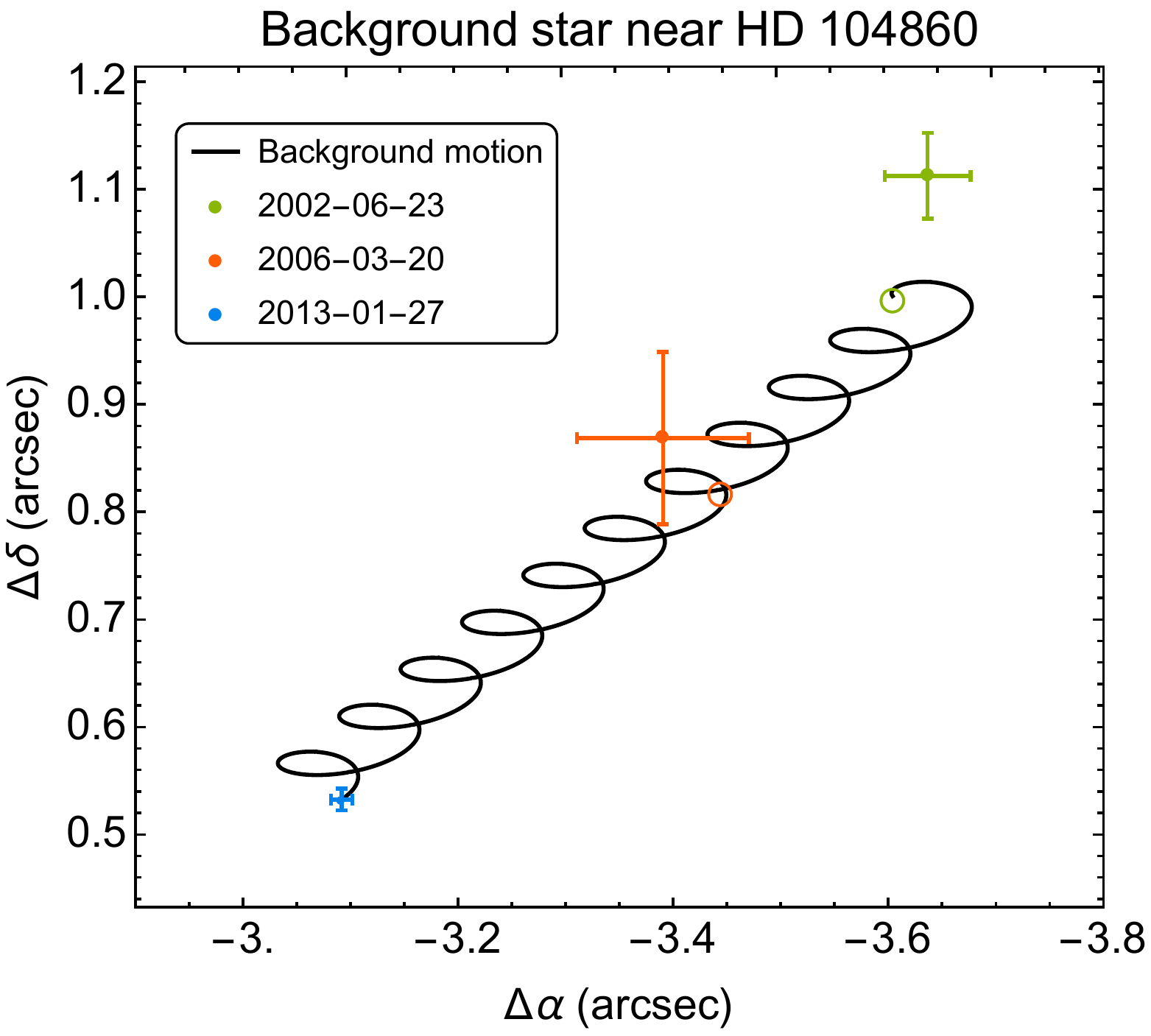}
\caption{Astrometry of the point source relative to \firstD, measured with Palomar-P1640 on 2002-06-23 (\modif{green}) \modif{from \citet{Metchev2009}}, HST-NICMOS on 2006-03-20 (\modif{orange}), and Keck-NIRC2 on 2013-01-27 (\modif{blue}). The \modif{black} line shows the relative motion of a fix background object from the point source position at epoch 2013-01-27. The \modif{empty circles} indicate the predicted background position at the corresponding epoch. The point source measured astrometry follows the background track and is not co-moving with \firstD.\label{fig:background}}
\end{figure}

We detect a bright point source $\sim3.5$\arcsec{} North-West from the star. We report its astrometry in Table~\ref{tab:background}. We identify it as candidate \#1 detected in \citet{Metchev2009}, and we confirm that its astrometry is consistent with a background star, as noted by the authors.
We note that this object is wrongly reported as a binary companion in the Washington Double Star catalog (WDS 12046+6620AB). To further confirm the background nature of this point source, we re-processed a Keck-NIRC2 archival dataset acquired on UT-2013-01-27 using the K' filter (program N117N2, PI: F. Morales). After standard processing and registration steps, we detect the point source using the classical Angular Differential Imaging \citep{Marois2006} PSF-subtraction method. We report the measured astrometry of the point source relative to \firstD in Table~\ref{tab:background}. The point source indeed follows the background track between the three epochs, although we note a significant shift between the measured position and the background-track-predicted astrometry over the 11-year baseline that is likely due to the background star's own proper motion (see Fig.~\ref{fig:background}).

\begin{deluxetable}{lrrc}
\tablecaption{Point source Astrometry around \firstD \label{tab:background}}
\tablehead{
 \colhead{Epoch}	& \colhead{Separation} & \colhead{P.A.} & \colhead{Ref.} \\
 \colhead{(UT date)}	& \colhead{(\arcsec)} & \colhead{(\degr)} & \colhead{} }
\startdata
2002-06-23	& $3.803\pm0.027$	& $-72.99\pm0.28$		&1\\ 
2006-03-20	& $3.50\pm0.08$	& $-75.6\pm1.2$		&2\\ 
2013-01-27	& $3.14\pm0.01$	& $-80.2\pm0.2$	&2\\
\enddata
\tablerefs{1: \citealt{Metchev2009}; 2: This work.}
\end{deluxetable}

\subsection{\secD}
\secD is a F0V star at $67\pm2$~pc  \citep{GaiaCollaboration2016}. \citet{Moor2006} reported a 50\% probability for the star to be a member of the IC 2391 supercluster, of age $50\pm5$~Myr \citep{BarradoyNavascues2004}. On the other hand, \citet{Chen2014} found the system to be a field star of isochronal age $\sim830$~Myr.
The system has a  fractional infrared excess around 100~$\mu$m with $L_{dust}/L_{\star}\sim 5.6\times 10^{-4}$ \citep{Moor2006}. It is well modeled by a two-temperature black-body emission, with a warm disk at $164\pm7$~K with a fractional luminosity of $\sim0.38\times 10^{-4}$ and a cold dust belt at $54\pm6$~K with a fractional luminosity of $\sim4.1\times 10^{-4}$ \citep{Chen2014}. Assuming disks in radiative and collisional equilibrium and composed of silicate spherical spheres, these black-body emission corresponds to dust belts of mass $\sim1.3\times 10^{-4}~M_{Moon}$ and $\sim8.7\times 10^{-1}~M_{Moon}$, respectively, and at radii  $\sim6$~au and $\sim152$~au, respectively.


We spatially resolve the outer disk in \secD in both the F110W and the F160W datasets. In the F110W image, the disk is detected up to S/N$\sim6$ per pixel, and S/N $\sim 45$ integrated over the pixels above 1.5$\sigma$. In F160W, the detection is found with S/N up to $\sim$ 4 per pixel, with S/N $\sim$ 21 integrated over the disk area above $1.5\sigma$.
The disk presents a similar morphology in the two images. They both reveal a disk of radius $\sim100$~au ($\sim1\farcs5$) and inclined by $\sim60$\degr from face-on. As for \firstD, the disk radius is significantly smaller than the disk radius estimated from SED modeling, which stresses again the need for imaging to put reliable constraints on disk properties. The South side of the disk is not detected, which also indicates anisotropic scattering by the dust grains, presumably showing the near-side of the disk North of the star.

\section{Disk modeling}\label{sec:analysis}

PSF subtraction with algorithms that solve the least-square problem of minimizing the residuals between the science image and a set of eigen-images systematically involves some level of over-subtraction of circumstellar materials, along with the PSF \citep{Lafreniere2007,Soummer2012,Pueyo2016}. This effect biases both the morphology and the photometry of circumstellar sources, and the effect depends on the shape of the source and of the reduction parameters. For extended objects in ALICE-processed NICMOS data, the algorithm throughput typically ranges from 20\% at 0\farcs5 from the star to 80\% at large separations, in the case of a face-on \modif{ring-like disk}.  

Calibrating the effect of the algorithm throughput is critical to properly characterize the morphology of extended sources, and to estimate their unbiased surface brightness. This is even more essential for inclined debris disks seen in scattered-light, since the algorithm throughput is lowest at short separations, along inclined disks' semi-minor axis where the effect of anisotropic scattering is the strongest \citep{Hedman2015,Perrin2015,Milli2017b}. Accounting for the post-processing throughput is thus required to accurately map a disk surface brightness distribution, measure its scattering phase function and flux density, and put constraints on its scattering properties.

In this section, we describe the parametric modeling used to constrain the disk morphologies, the forward modeling method used to calibrate the post-processing throughput, and the models that best fit the disks detected around \firstD and \secD.

\subsection{Forward modeling method}

Assuming that the morphology of the astrophysical source is known, the post-processing throughput can be inferred though forward modeling. 
To constrain the morphology and photometry of the disks detected around \firstD and \secD, we used the same methodology as in \citet{Choquet2016}, using parametric modeling and the analytical forward modeling method \modif{\citep[valid with PCA-type algorithms,][]{Soummer2012,Pueyo2016}. This method consists in subtracting from the source's model its projection on the eigen-vectors used to process the data.} 

We generated a grid of disk models with a set of free parameters that we wish to constrain, and convolved each model by a synthetic unocculted NICMOS PSF generated in the corresponding filter with the Tiny TIM package \citep{Krist2011}. From each model, we then subtracted its projection on the same eigen-vectors as used for the science data, intrinsically using the same reduction zone and number of PCs. This process removes from the raw input model the part over-subtracted during post-processing, and reveals the (hereafter) forward model. 
We then rotated and combined these forward models with the same parameters and angles as the science images, and we scaled the resulting model to the total flux of the disk in the science image. Each forward model is then directly compared to the disk image.

We used the GRaTer radiative transfer code to create the disk models \citep{Augereau1999a,Lebreton2012}. This code computes optically thin centro-symmetric disk models assuming a \citet{Henyey1941} scattering phase function of asymmetric parameter $g$. The dust density distribution $n(r,z)$ is parametrized with a radial profile $R(r)$ falling off with power laws $\alpha_{in}$ and $\alpha_{out}$ inward and outward from a parent belt at radius $R_0$, and with a Gaussian vertical density $Z(r,z)$ profile with a scale height $\zeta(r)$ rising linearly with an aspect ratio $h$: 
\begin{flalign}
n(r,z)\propto R(r) Z(r,z), &&
\end{flalign}
with:
\begin{flalign}
R(r) =\left( \left( \frac{r}{R_0}\right)^{-2\alpha_{in}} +\left( \frac{r}{R_0}\right)^{-2\alpha_{out}} \right)^{-1/2}&&
\end{flalign}
\begin{flalign}
Z(r,z) =\exp \left( - \left( \frac{|z|}{\zeta(r)}\right)^{2} \right) &&
\end{flalign}
\begin{flalign}
\zeta(r) =h r &&
\end{flalign}
The disk center relative to the star's position is parametrized by offsets $du$ and $dv$ in the disk plane, with $du$ along the major axis (unaffected by projections effect), and $dv$ along the perpendicular axis (appearing projected along the disk's minor axis). The model image is simulated with a position angle (PA) parametrized by $\theta$, and inclination~$i$. 
Given the geometry of the two systems, the vertical scale heights of the disks are not properly constrained by our data. We fix the aspect ratio to $h=0.05$, a reasonable thickening assumption for unperturbed debris disks due to the combined action of radiation pressure and grains mutual collisions \citep{Thebault2009}. Furthermore, given the inclinations of the disks, a Henyey-Greenstein scattering phase function (SPF) model will describe the surface brightness variations in the disks over a  range of scattering angles limited by the disk inclinations, but may not properly describe the actual SPF over all angles \citep[see][]{Hedman2015}.

We estimated the goodness of fit of our models to the disk image by computing the reduced chi square $\chi^2_{red}$ value over a large area encompassing the disk  in the image.  After identifying the best model within the grid, we  refine the best-fit parameters  by interpolating the reduced chi square values around the best model in the grid, independently for each parameter. The best chi square values are  larger than 1 because the noise maps are overall slightly underestimated, although very representative of the noise spatial distribution (e.g. the noise signature from spider residuals). To estimated the uncertainties on the model parameters, we thus normalize the chi square by its best value, assuming that the noise underestimation is common to all pixels. We then estimated the uncertainty on each parameter assuming that our estimator follows a chi-squared distribution, from the values at which the interpolated $\chi^2_{red}$ increased by $1\sigma=\sqrt{2/N_{dof}}$ from 1, with $N_{dof}$ the number of degrees of freedom in the fit. 

\subsection{\firstD analysis}

\begin{deluxetable}{lrrr|rr}
\tablecaption{Parameter grid and best model for \firstD  \label{tab:grid104}}
\tablehead{
 \colhead{Param.}	& \colhead{Min.}	&\colhead{Max.}	& \colhead{$N_{val}$} 	& \colhead{Best Model}	& \colhead{Best Model}\\
  \colhead{}		& \colhead{}		&\colhead{}		& \colhead{} 			& \colhead{(in grid)}		& \colhead{(interpolated)\tablenotemark{a}}
 }
\startdata
$R_0$ (au)           		& 106 	&122	    	&5	&114		&$114\pm6$	\\
$|g|$                     		&0.0       	&0.4     	&5	&0.2		& $0.17\pm0.13$\\
$\theta$ (\degr)			&-5		&7		&5	&1		&$1\pm5$		\\
$i$ (\degr)				&52		&64		&5	&58		&$58\pm5$	\\
$\alpha_{in}$ 			&2	        &10		&5	&10		&$\ge4.5$		\\
$\alpha_{out}$ 			&-6	        &-2		&5	&-4		&$-3.9\pm1.6$		\\
$du$ (au)				&-10		&10		&5	&0		&$2\pm7$		\\
$dv$ (au)				&-20		&10		&7	&-5		&$-7\pm13$	\\
$\chi^2_{red}$			&\nodata	&\nodata	&\nodata	&1.833	&1.826		\\
\enddata
\tablenotetext{a}{Shows $1\sigma$ uncertainties.}
\end{deluxetable} 

\begin{figure}
\center
\includegraphics[width=8.5cm]{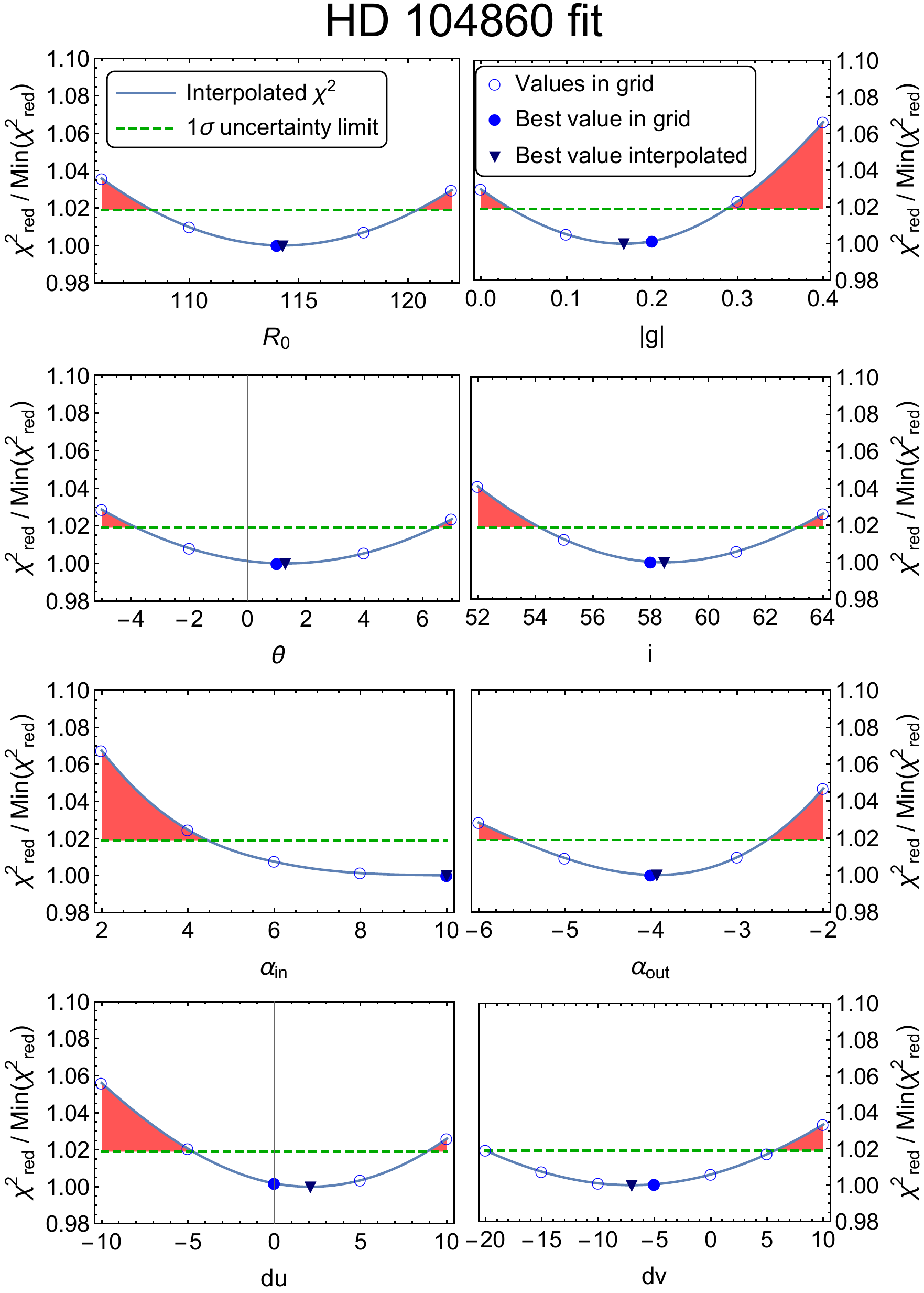}
\caption{Goodness of fit for each parameter modeling the disk around \firstD. The chi square values shown for each parameter value in the grid (empty circles) have all the other parameters fixed to their best values in the grid (filled circles). The $\chi^2$ is interpolated between each parameter value in the grid to refine the best fit values (filled triangles). The dashed green line shows the $1\sigma$ threshold used to estimate the uncertainties on the parameters, and the red-shaded areas show the parameters values ruled out by our modeling.   \label{fig:fitHD104}}
\end{figure}

\begin{figure*}
\center
\includegraphics[width=18cm]{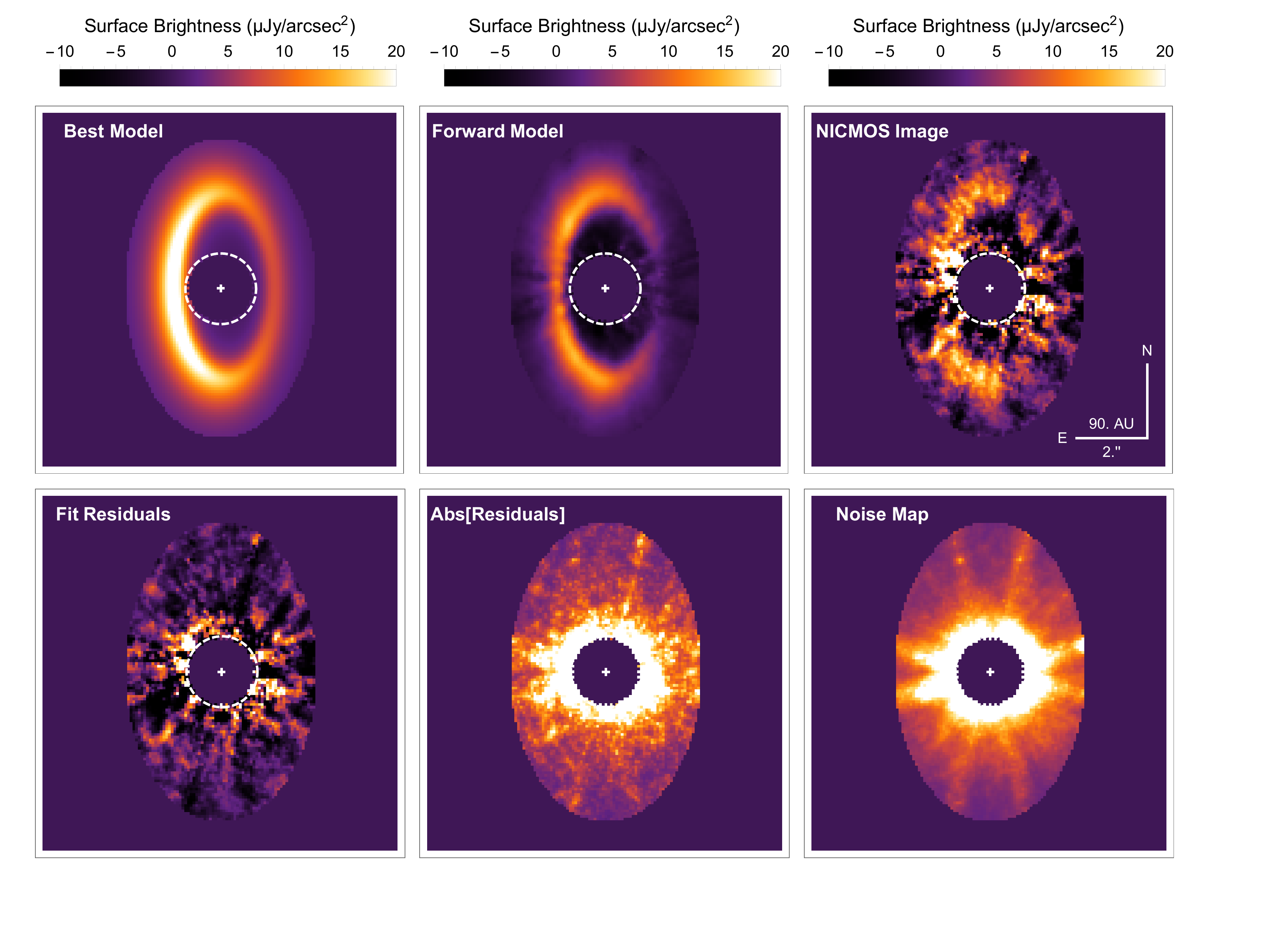}
\caption{Best model of the debris disk detected around \firstD. The top row shows The best raw model convolved by a NICMOS PSF (left), the model affected by over-subtraction after forward modeling (middle), and the NICMOS image for comparison (right). The bottom row shows the residuals after subtracting the forward model from the NICMOS image (left), the absolute value of these residuals (middle), and the noise map for comparison (right). No disk structure is appearing in the residual map, indicating that the model properly fit the data. All images are displayed within the elliptical mask used for the fit, and have been smoothed by convolution with a synthetic PSF. \modif{The raw model was hence convolved twice with a synthetic PSF, once to account for the diffraction and once for smoothing and comparison with the other smoothed images.}\label{fig:HD104Model}}
\end{figure*}

\begin{figure}
\center
\includegraphics[width=8.5cm]{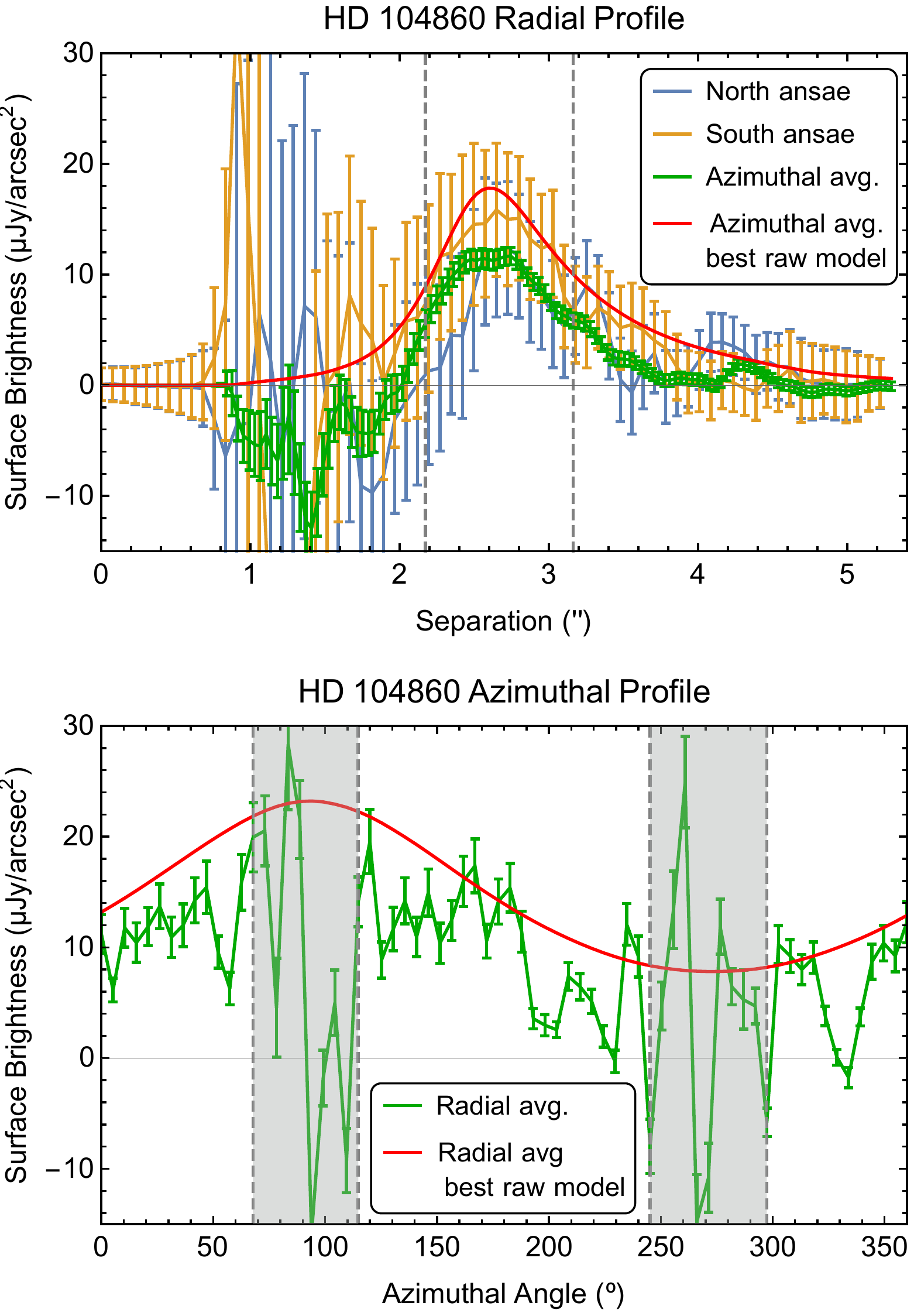}
\caption{Radial and azimuthal profiles of \firstD, measured after de-projecting the disk from its inclination, position angle and offset from the star. Top:  the blue and orange lines show the disk profile in the North and South ansae respectively. The green line show the radial profile averaged regardless of anisotropy of scattering over all azimuthal angles (excluding the angles around 90\degr{} and 270\degr{} delimited by the gray dashed lines in the bottom plot). All three profiles are affected by over-subtraction induced by post-processing. The red line shows the average radial profile measured in the best raw model, before forward modeling, and is representative of the disk surface brightness unaffected by over-subtraction artifacts. The dashed gray lines show the disks full-width at half-maximum. 
Bottom: Azimuthal profile of the disk average over the disk full-width at half-maximum (green line). The red line shows the average azimuthal profile of the best model before forward modeling, and corresponds to a Henyey-Greenstein scattering phase function of parameter $g=0.17$. The dashed gray lines show azimuthal angles  where the disk is dominated by PSF residuals, along the minor axis at 90\degr{} and 270\degr{}. The 1$\sigma$ uncertainties are computed accordingly from the noise map.\label{fig:HD104profiles}}
\end{figure}

\begin{figure}
\center
\includegraphics[width=8.5cm]{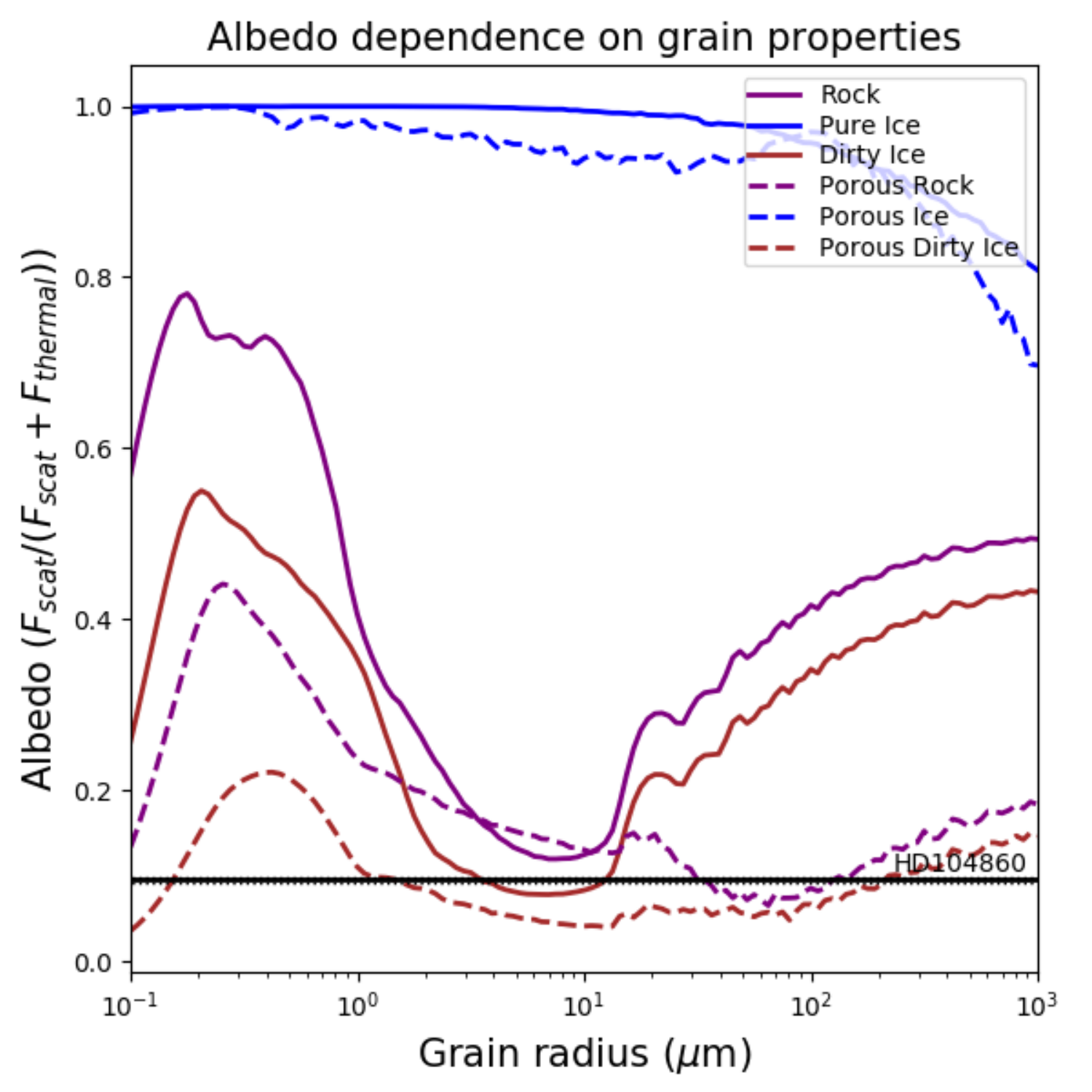}
\caption{Scattering albedo computed under the Mie theory as a function of grain size for a disk with \firstD's best fit morphology, assuming different grain compositions: pure ice (blue), dirty ice (red), silicates (purple), and different porosities: compact grains (solid lines) and 90\% porous grains (dashed lines). The measured scattering albedo for \firstD (black line) rules out water ice compositions and is consistent dirty ice grains larger than $\sim2~\mu$m. \label{fig:HD104albedo}}
\end{figure}

For \firstD, we generated a grid of 546875 models with 8 free parameters to constrain its morphology. Table~\ref{tab:grid104} describes the simulated parameter ranges and best fit values. The goodness of fit of each parameter is also presented in Fig.~\ref{fig:fitHD104}. The data were fit within an elliptical area of semi-major axis 4\farcs2 oriented North-South and semi-minor axis 2\farcs6, excluding the central area masked during post-processing, which resulted in $N_{dof}=5536$ degrees of freedom in the fit. This area excludes the background star at $\sim3.5\arcsec$ from the star.  The image of the best model and comparison to the NICMOS image and noise map are presented in Fig.~\ref{fig:HD104Model} within the fit area.

We find that the best fit to \firstD's data is a disk of radius $R_0=114\pm6$~au, inclined by $i=58\pm5\degr$ from face-on, with a position angle of $\theta=1\pm5\degr$ East of North. These values are consistent with the morphology of the disk observed at 100~$\mu$m and 160~$\mu$m by \citet{Morales2013}. We find a relatively low value for the Henyey-Greenstein parameter of anisotropic scattering, with $|g|=0.17\pm0.13$ indicative of grains favoring forward scattering. This value is expected for a disk of moderate inclination such as \firstD, which does not probe very small scattering angles ($> 32\degr$):  slightly forward scattering SPFs with Henyey-Greenstein parameter around $|g|$=0.1--0.3 have been observed for many debris disks with similar or lower inclinations \citep[HD 92945, HD107146, HD 141569, HD207129, Fomalhaut][]{Golimowski2011, Ardila2004,Mawet2017a,Krist2010,Kalas2005a}. As mentioned before, this may not be representative of the scattering phase function at smaller scattering angles, as it was shown that, for some dust grains, it may significantly differ from a Henyey-Greenstein model and sharply peak at angles below 40\degr{} despite a relatively flat phase function at large angles \citep{Hedman2015,Milli2017b}.

Our analysis show that the disk has a ring shape with an asymmetric radial density profile. We find that the disk outer edge follows a power law in $\alpha_{out}=-3.9\pm1.6$, corresponding to a surface density in $r^{\Gamma_{out}}$ with $\Gamma_{out}=\alpha_{out}+1=-2.9\pm1.6$, assuming a scale height rising linearly with radius \citep{Augereau1999a}. This is consistent within our uncertainties with disk evolution models, which predict an outer surface density profile in $\Gamma_{out}=-1.5$ for small grains created under steady state collisions and set on eccentric orbits by radiative pressure, accumulating in the outskirts of their birth ring \citep{Strubbe2006,Thebault2008}. We find that the disk inner edge is very sharp, with a lower limit of $\alpha_{in}\ge4.5$ on the inward power law of its radial density  profile. Sharp inner edges in debris disk can be sculpted by planets orbiting within the disk, confining the dust out of a chaotic zone through mean motion resonances \citep{Wisdom1980,Mustill2012}, while unperturbed systems have smoother inner edges filled by small grains due to Poyting-Robertson drag. We do not find significant offsets of the ring with respect to the star within $\pm7$~au along the disk major axis, and within $\pm13$~au projected on the minor axis, indicating that a planet responsible for carving the disk inner edge would have a low eccentricity. 

Using the best fit morphological parameter values, we de-project the NIMCOS image based on the disk inclination, position angle, and offsets, and compute its radial and azimuthal average profiles (see Fig.~\ref{fig:HD104profiles}). In the following, we quantify the disk photometry  using both the NICMOS image, which is affected by over-subtraction as shown in Fig.~\ref{fig:HD104Model}, and  the best  model before forward modeling, which is free of post-processing artifacts but is entirely model-dependent.

In the NICMOS image, we measure a surface brightness on the disk spine of $11.7\pm0.8~\mu$Jy/arcsec$^2$ averaged over all scattering angles excluding $91\pm23\degr$ and $271\pm26$\degr, where stellar residuals from the telescope spider dominate the disk brightness. We find a corresponding average surface brightness of $S=17.8~\mu$Jy/arcsec$^2$ in the best model unbiased by over-subtraction. The South ansae is $\sim30$\% brighter than the North one in the NICMOS image ($S_{S}=16\pm6~\mu$Jy/arcsec$^2$ and $S_{N}=12\pm5~\mu$Jy/arcsec$^2$ respectively), but the asymmetry is not significant given our uncertainties. From the stellar flux ($F_\star=3.1$~Jy  in the NICMOS F110W filter), we estimate that the disk has a typical reflectance of $R=S/F_\star=(5.7\pm0.3)\times10^{-6}$~arcsec$^{-2}$, based on the mean surface brightness of the model. We estimate the disk flux density to $F_{scat}=200\pm5~\mu$Jy, integrated in the best model over an elliptical surface of 20~arcsec$^2$ with outer semi-major axis 3\farcs9, inner semi-major axis 1\farcs6, inner and outer semi-minor axis projected from the semi-major axis values by the best fit inclination $\cos(i)$, and with a position angle equal to the best fit value $\theta$. The flux density integrated in the NICMOS image over the same surface is $92\pm5~\mu$Jy, significantly affected by the post-processing throughput. Compared with the flux of the star, this correspond to a scattering efficiency of $f_{scat}=F_{scat}/F_\star=(67\pm2)\times 10^{-6}$, based on the model's flux density.

The scattering efficiency provides a measure of the dust scattering albedo $\omega$ knowing the total luminosity received by the grains \citep{Krist2010,Golimowski2011}:
\begin{flalign}
\omega=\frac{f_{scat}}{f_{emit}+f_{scat}}, &&\label{eq:albedo}
\end{flalign}
with $f_{emit}=L_{dust}/L_\star$ the disk infrared fractional luminosity. The scattering albedo is an empirical quantity that provides a degenerate, wavelength-averaged combination of the grains albedo and of the disk scattering phase function integrated over the scattering angles probed by the disk geometry. The true albedo of the dust can only be recovered with assumptions on the disk's SPF.  
Assuming that our scattering efficiency measurement integrates the light scattered by the grains responsible for the thermal emission, and using the infrared fractional luminosity of $f_{emit}=6.4\times 10^{-4}$ reported for the outer disk by \citet{Morales2016} (well-constrained from 24~$\mu$m  to 1.3~mm  photometry), we find a scattering albedo of $\omega=9.5\pm0.3$\%. 

We compared this value to the scattering albedo of grains with various compositions and porosities, computed under the Mie theory for a disk with the best-fit morphology found for \firstD in the F110W filter (Fig.~\ref{fig:HD104albedo}). We find that our scattering albedo estimation rules out pure water ice composition and is consistent with dirty ice grains \citep{Preibisch1993} larger than $\sim3~\mu$m in the case of compact grains, and larger than $\sim1~\mu$m in the case of 90\% porous grains. Assuming compact grains, this is consistent with particles larger than the blowout size as suggested by \citet{Pawellek2014} for this disk (typical grain size of 7~$\mu$m, blowout size of 0.4~$\mu$m), although they used pure silicate grains in their study. Assuming 90\% porous grains, the blowout size is 10 times larger than for compact grains which is consistent with our scattering albedo value. A few other debris disks,  e.g. HD 181327 and HD 32297, are suspected to have porous grains \citep{Lebreton2012,Donaldson2013}. However, they are also much brighter than \firstD, which may indicate also a different composition.

\subsection{\secD analysis}

\begin{deluxetable*}{lrrr|rr|rr}
\tablecaption{Parameter grid and best models for \secD  \label{tab:grid192}}
\tablehead{
  \colhead{}		& \colhead{}		&\colhead{}		& \colhead{} 			& \colhead{F110W}		& \colhead{F110W}	& \colhead{F160W}		& \colhead{F160W}\\
 \colhead{Param.}	& \colhead{Min.}	&\colhead{Max.}	& \colhead{$N_{val}$} 	& \colhead{Best Model}	& \colhead{Best Model}& \colhead{Best Model}	& \colhead{Best Model}\\
  \colhead{}		& \colhead{}		&\colhead{}		& \colhead{} 			& \colhead{(in grid)}		& \colhead{(interpolated)\tablenotemark{a}}& \colhead{(in grid)}		& \colhead{(interpolated)\tablenotemark{a}}
 }
\startdata
$R_0$ (au)           		& 81 		&109	    	&5		&95		&$95\pm12$		&95		&$95\pm9$	\\
$|g|$                     		&0.15       	&0.6     	&10  	   	&0.3		&$0.29\pm0.12$	&0.4		&$0.41\pm0.16$	\\
$\theta$ (\degr)			&-101	&-77		&7	     	&-93		&$-93\pm7$		&-85		&$-85\pm8$	\\
$i$ (\degr)				&50		&70		&6		&62		&$60\pm8$		&58		&$58\pm7$	\\
$\alpha_{in}$ 			&1	        &11		&6		&7		&$\ge1.4$			&11		&$\ge2.9$		\\
$\alpha_{out}$ 			&-3.5	&-1		&6		&-2		&$-2.0\pm0.8$		&-2		&$-2.0\pm0.9$	\\
$du$ (au)				&-10		&25		&8		&10		&$11\pm15$		&5		&$4\pm12$	\\
$\chi^2_{red}$		&\nodata	&\nodata	&\nodata		&4.864	&4.857			&2.247	&2.246	\\
\enddata
\tablenotetext{a}{Shows $1\sigma$ uncertainties.}
\end{deluxetable*} 

\begin{figure}
\center
\includegraphics[width=8.5cm]{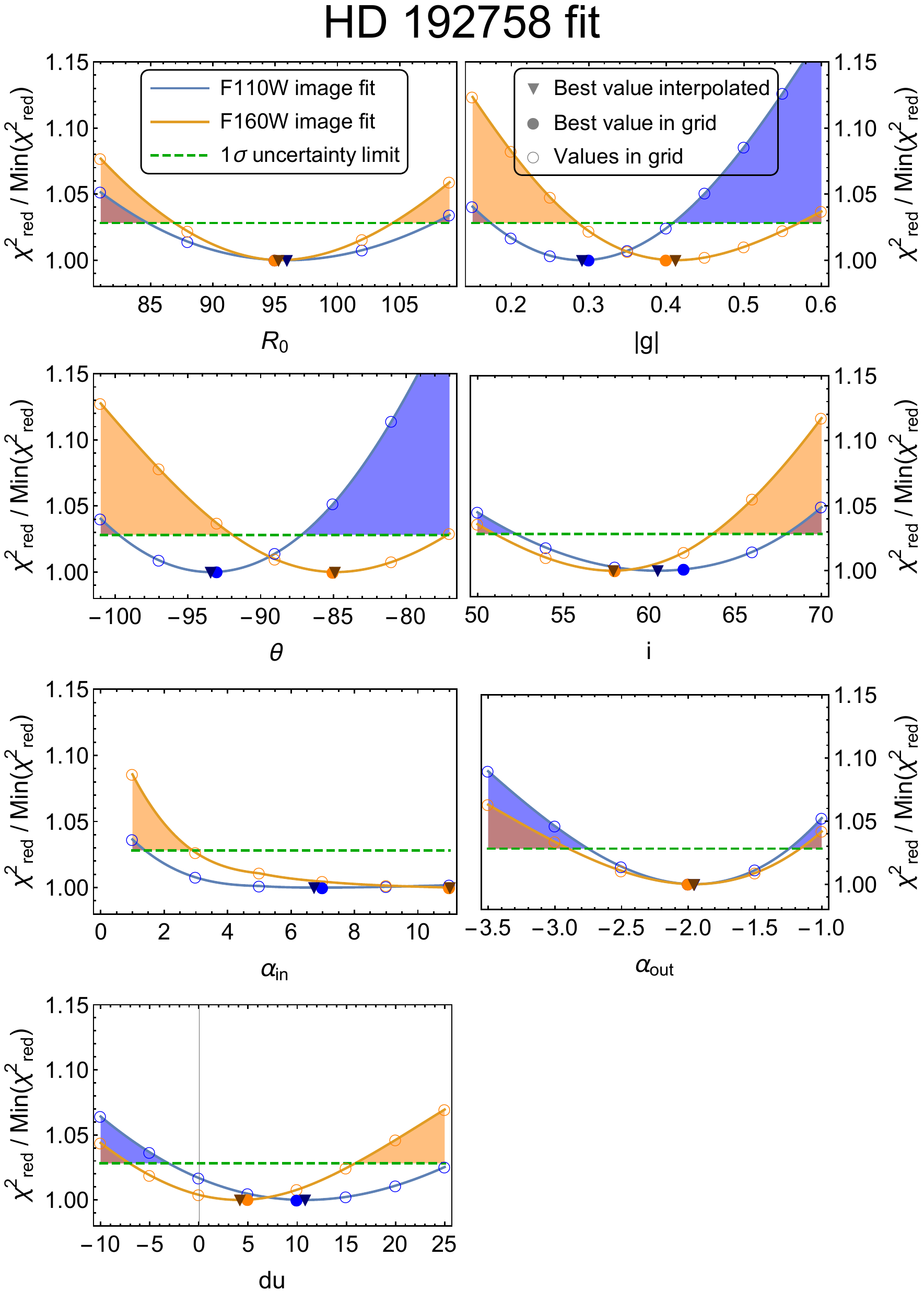}
\caption{Goodness of fit for each parameter for the disk around \secD, in blue for the F110W filter dataset, and in orange for the F160W dataset. The chi square values shown for each value in the grid (empty circles) have all the other parameters fixed to their best values in the grid (filled circles). The $\chi^2$ is interpolated between each parameter value in the grid to refine the best fit values (filled triangles). The dashed green line shows the $1\sigma$ threshold used to estimate the uncertainties.The shaded areas  show the parameters values ruled out by our modeling.   \label{fig:fitHD192}}
\end{figure}

\begin{figure*}
\center
\includegraphics[width=18cm]{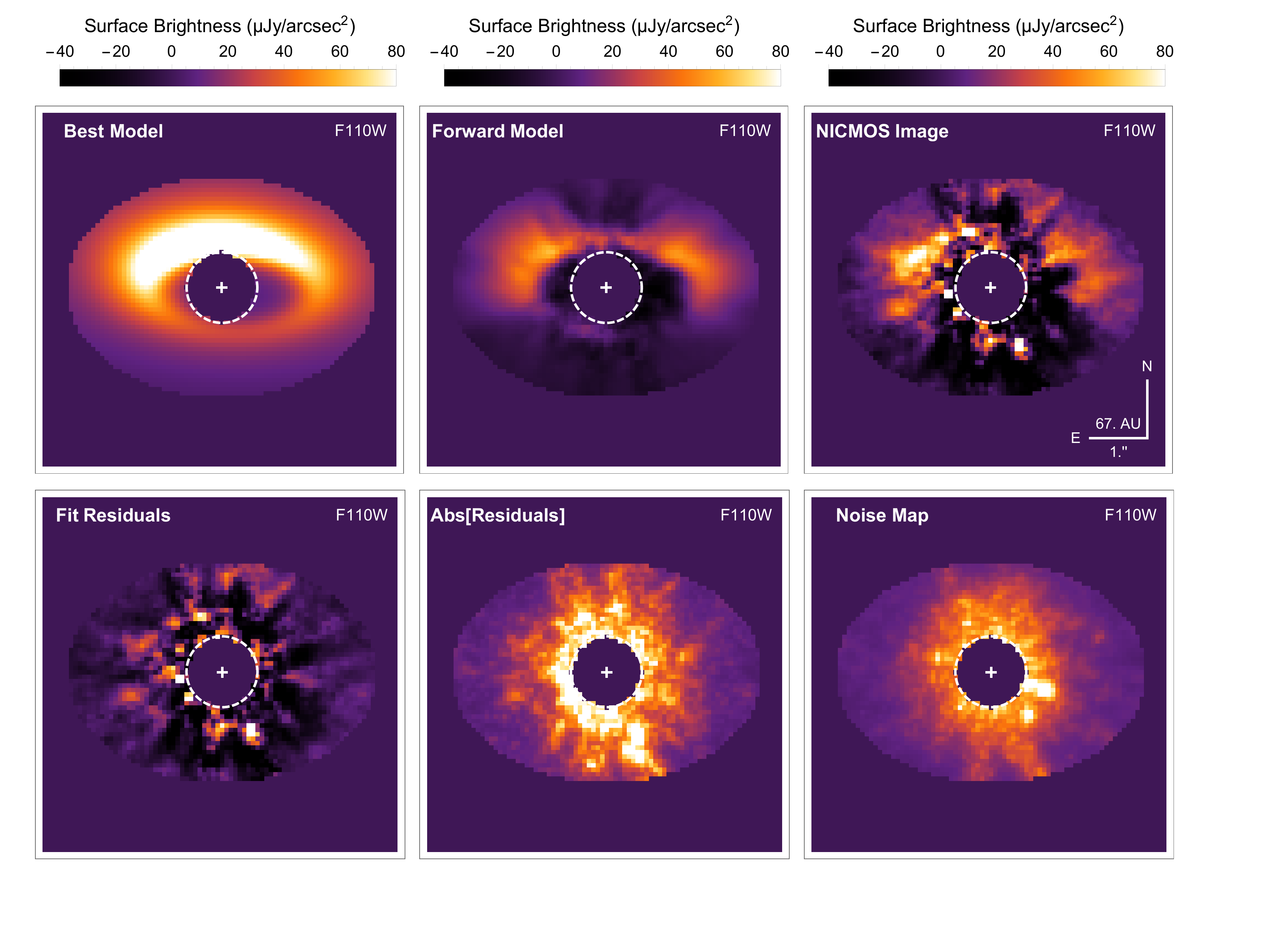}
\caption{Same as Fig.~\ref{fig:HD104Model} for \secD in the F110W filter.\label{fig:HD192Model_F110W}}
\end{figure*}

\begin{figure*}
\center
\includegraphics[width=18cm]{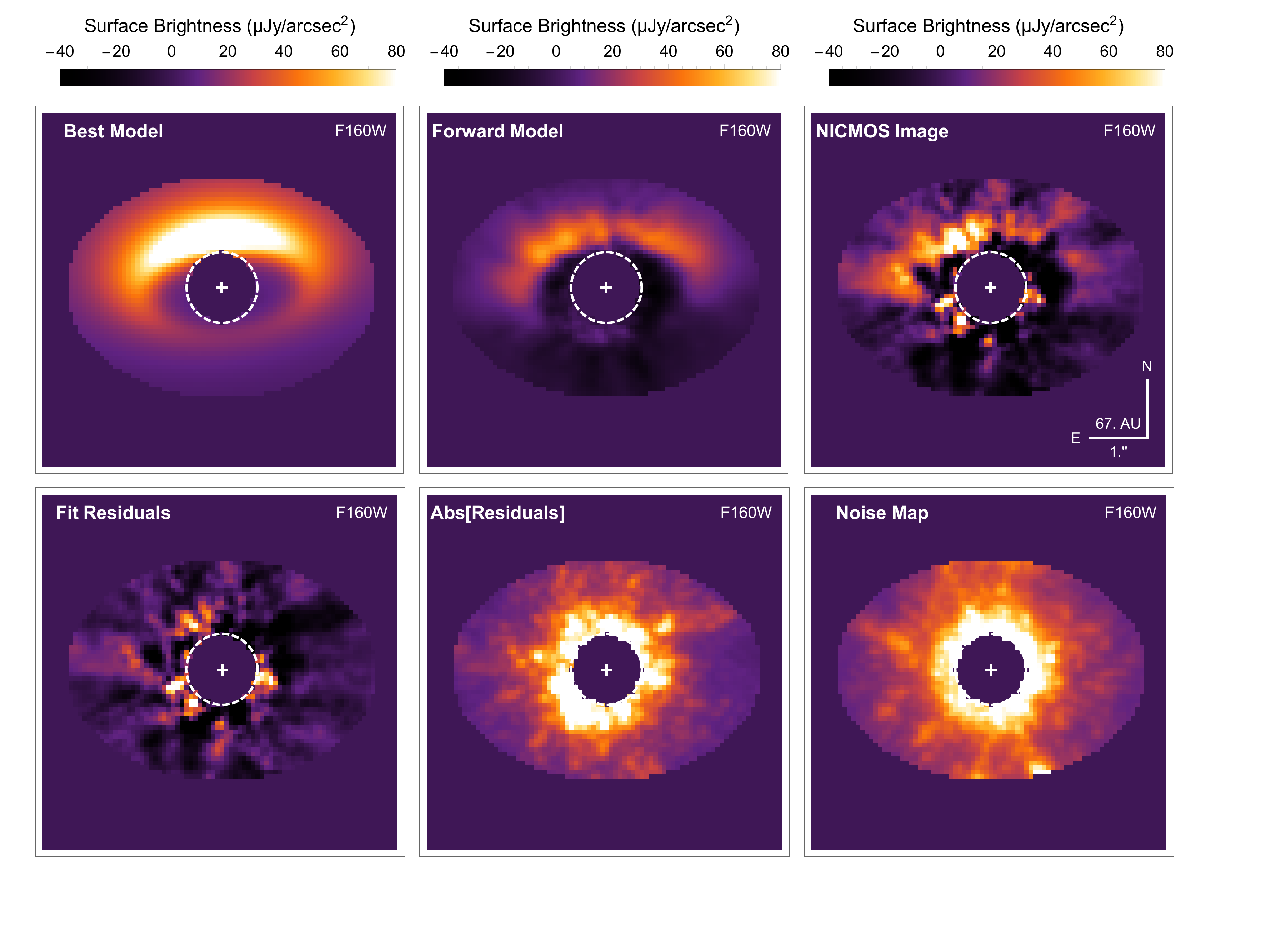}
\caption{Same as Fig.~\ref{fig:HD104Model} for \secD in the F160W filter.\label{fig:HD192Model_F160W}}
\end{figure*}

For \secD we generated a single grid of models that we fit separately to the  F110W and F160W images. It is indeed reasonable to assume a similar dust density distribution in both images but possibly different scattering properties at the two different wavelengths (albedo, scattering phase function). 
We simulated 604800 models with 7 free parameters, fixing the offset along the disk minor-axis to  $dv=0$, as this parameter is unconstrained by our data given the disk geometry. We present the simulated parameter ranges and best fit values  in Table~\ref{tab:grid192} for both filters. The goodnesses of fit  are presented jointly in Fig.~\ref{fig:fitHD192}. The shaded areas show better constraints on some parameters when combining the fits to both datasets. We computed the chi square values in an elliptical area of semi-major axis 2\farcs6 oriented East-West, and semi-minor axis 1\farcs9, and excluding the central area masked during post-processing. The same area was used for both datasets, and corresponds to $N_{dof}=2525$ degrees of freedom. We show in Fig.~\ref{fig:HD192Model_F110W} the best fit to the F110W dataset and in Fig.~\ref{fig:HD192Model_F160W} the best fit to the F160W data, along with their respective comparison to the NICMOS images and noise maps.

We find that the best fit parameters are consistent for both datasets. In the following, we describe the best model by averaging the F110W and F160W best parameter values and the uncertainties combined from the two fits, which provides better constraints, except for two parameters ($|g|$ and $\theta$) as discussed below. The disk has a radius of $R_0=95\pm9$~au and is seen inclined by $i=59\pm7$\degr{} from face-on. The best fit PA differs by 8\degr{} between the two datasets. This may be due to starlight residuals from the telescope spider, at a comparable orientation to the disk major axis, which may bias the disk position angle. The mean PA between both fits is $\theta=-89\pm12$\degr{} East of North, using conservative error bars encompassing uncertainties form the two fits. The disk has a relatively low Henyey-Greenstein parameter of anisotropic scattering, with  values of $|g|=0.29\pm0.12$ in the F110W image, and $|g|=0.41\pm0.16$ in the F160W image. Differing values can be explained by different scattering properties at different wavelengths, although we do not expect a significant difference, as the two bandpasses are relatively close. \modif{As for \firstD}, these relatively low values are consistent with the finding of \citet{Hedman2015} \modif{about the apparently low $g$ values of inclined disks observed only near scattering angles of 90\degr{}}. 

We find that the dust density follows a radial power law in $\alpha_{out}=-2\pm0.8$ outward from the parent radius, which is very consistent with the sharpness expected from evolution models with small grains on eccentric orbits accumulating outside of the main belt colliding zone \citep{Thebault2008}. Given the disk inclination and the quality of our images, we can only put a weak constraint on the radial profile inward from the parent radius $R_0$, with a power law coefficient steeper than $\alpha_{in}>2.9$. We find that the disk center may be offset by $du=7\pm12$~au along the major axis but we cannot rule out a system centered with the star.

As shown in Fig.~\ref{fig:HD192Model_F110W} and \ref{fig:HD192Model_F160W}, over-subtraction from post-processing strongly affects the disk photometry, due to its compact appearance in the image (1\farcs4 semi-major axis, and 0\farcs7 semi-minor axis at the edge of the reduction mask). The averaged radial and azimuthal profiles measured in the NICMOS images are thus  noisy and biased by over-subtraction so we relate hereafter photometric values from the best models before forward modeling only. In the F110W filter, we measure a mean surface brightness of $S^{F110W}=108\pm3~\mu$Jy/arcsec$^2$ in the North ansae, and of $S^{F160W}=81\pm5~\mu$Jy/arcsec$^2$ in the F160W data. Given the stellar flux in these bands ($F_{\star}^{F110W}=4.85$~Jy and $F_{\star}^{F160W}=3.26$~Jy respectively), the disk reflectances in the North ansae are $R^{F110W}=(22.3\pm0.6)\times10^{-6}$~arcsec$^{-2}$ and $R^{F160W}=(25\pm2)\times10^{-6}$~arcsec$^{-2}$ in the F110W and F160W NICMOS filters. 
By integrating the disk flux over a 7.1~arcsec$^2$ elliptical area (outer and inner semi-major axes 2\farcs3 and 1\farcs0, oriented with the best PA $\theta$ east of north, minor axes projected by $\cos(i)$ from the major axes values), we find flux densities of $F_{scat}^{F110W}=411\pm9~\mu$Jy in the F110W filter and $F_{scat}^{F160W}=319\pm12~\mu$Jy in the F160W filter. After normalizing by the stellar contribution, we find that the disk around \secD has  scattering efficiencies of $f_{scat}^{F110W}=(85\pm2)\times10^{-6}$ and $f_{scat}^{F160W}=(98\pm4)\times10^{-6}$, respectively in the two NICMOS filters.

From these scattering efficiency measurements in two different NICMOS filters, we can measure the disk color. We find  it has a scattering efficiency ratio of $f_{scat}^{F110W}/f_{scat}^{F160W}=0.87\pm0.05$ (F110W-F160W  color index of 0.15), indicative of red grains in the disk. Assuming spherical silicates grains following the Mie theory, this color is consistent with the lack of particles smaller than $\sim$0.5~$\mu$m in the  parent belt, which is also consistent with the minimum size $a_{blow}=1.4~\mu$m under which silicate grains are blown out from the system by radiative pressure, assuming a stellar luminosity $L=4.9~L_\sun$, stellar mass of $M=1.2~M_\sun$, and a dust mass density of $\rho=3.3$~g.cm$^{-3}$.

\begin{figure}
\center
\includegraphics[width=8.5cm]{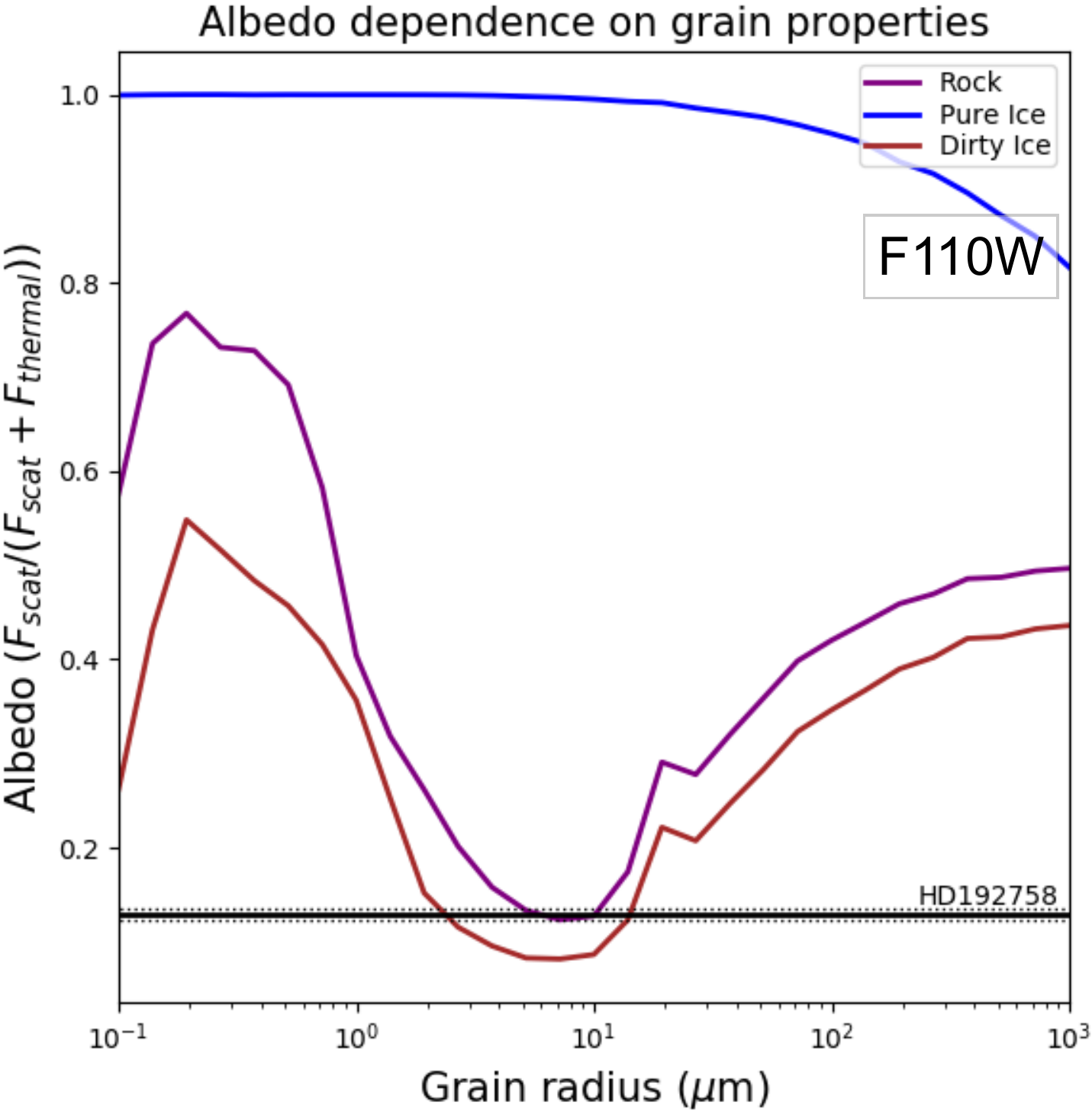}
\includegraphics[width=8.5cm]{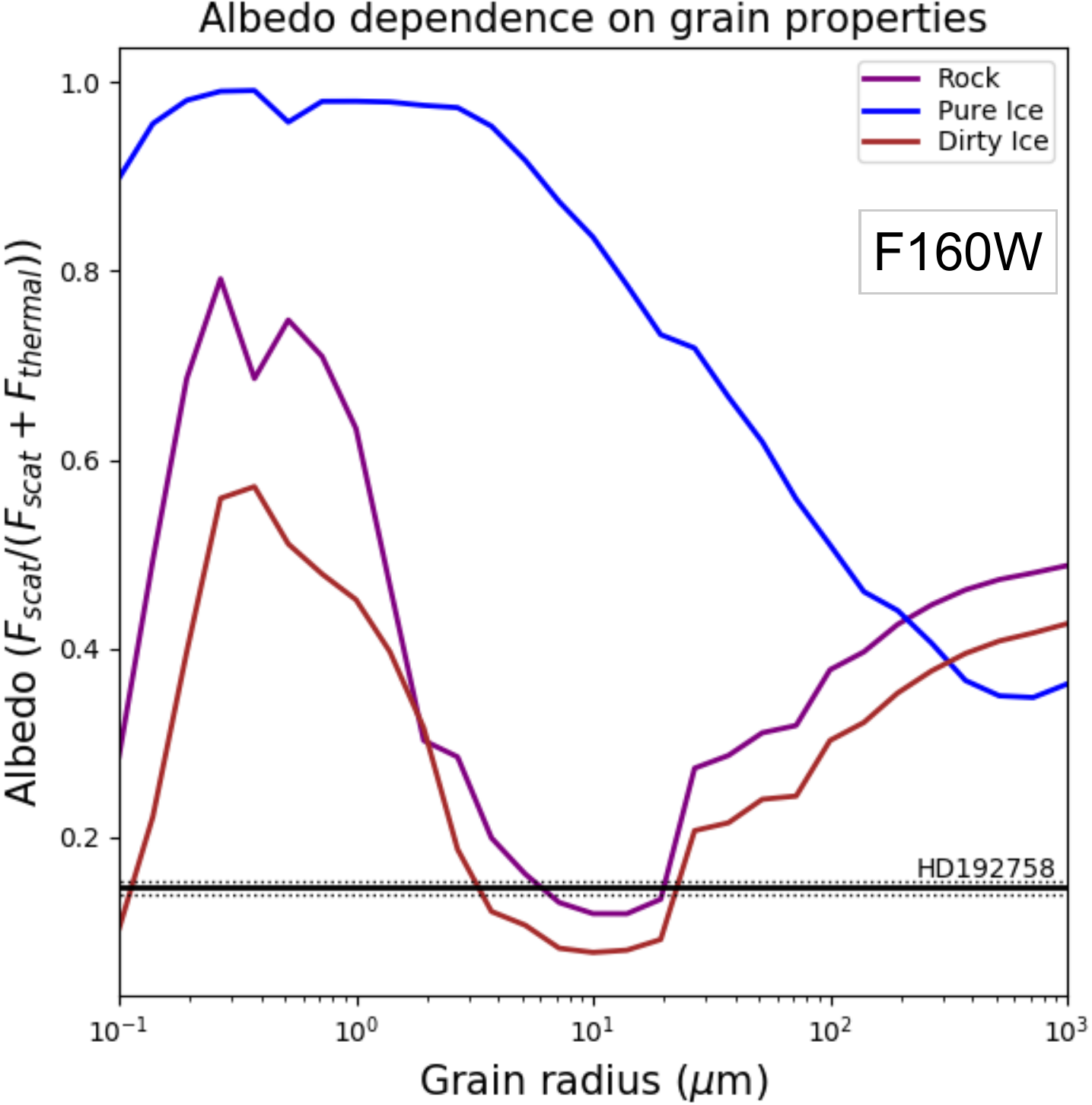}
\caption{Scattering albedo as a function of grain size for a disk with \secD's best fit morphology, assuming different grain compositions: pure ice (blue), dirty ice (purple), silicates (red), and computed in the F110W filter (top) and F160W filter (bottom). The measured scattering albedo for \secD (black line) rules out pure water ice compositions and is consistent dirty ice grains larger than $\sim3~\mu$m and pure silicate grains larger than $\sim6~\mu$m. \label{fig:HD192albedo}}
\end{figure}

Based on the disk infrared fractional luminosity $f_{emit}=(5.7\pm0.3)\times10^{-4}$ from \citet{Moor2011a}, we find scattering albedo values of $\omega^{F110W}=13.0\pm0.6$\% and $\omega^{F160W}=14.7\pm0.7$\% in the F110W and F160W filters, respectively. As found for \firstD, these values rule out pure water ice composition (see Fig.~\ref{fig:HD192albedo}). They are consistent with compact silicate grains larger than $\sim6~\mu$m, which is significantly larger than the blowout size $a_{blow}=1.4~\mu$m of for this system, as well as with dirty ice grains larger than $\sim3~\mu$m.

\section{Discussion}

These two detections add  to a growing population of debris disks resolved in scattered light. To date, 41 of such have been imaged around stars from $\sim10$~Myr to a few Gyr, over a large range of spectral types (See Fig.~\ref{fig:alldisks}). Yet, considering the numerous attempts to image debris disks around systems with large infrared excess indicative of a massive dust belt presumably detectable with basic geometric and albedo assumptions, this remains a relatively low number of detections. Three reasons could explain non-detections, given that the these surveys were designed to reach  surface brightness limits based on assumptions on the disk radii and albedos. 
\begin{enumerate}
\item \emph{Inaccurate radius estimations}: radii estimations solely based on black-body fit of SEDs are known to be degenerate with the dust size distribution and can be very inaccurate. The missed disks may have radii different from those inferred from SED modeling by a factor 0.5 to 2. They could  either be too compact to be observable with the $\sim 0.3$\arcsec{} typical inner working angle (IWA) of current coronagraphic instruments, or be larger and thus fainter than expected, falling short of the sensitivity limits planned for these surveys.   
\item \emph{Simplistic SPF assumptions:} exposure times assuming isotropic scattering may have been under-estimated if the disks have more complex scattering phase functions, depending on their inclination to our line of sight. \modif{Recent work showed that the minimum grain sizes estimated by SED modeling are often several microns and much larger than the blowout size \citep{Pawellek2015}, which suggests that wrong estimations may have been used on the grain size distribution and scattering phase function in older surveys.}
Furthermore, low-inclination disks not only appear fainter than expected if their SPF peaks at unprobed scattering angles, but they are also harder to detect with most post-processing techniques: they are self-subtracted with ADI-based algorithms, and over-subtracted with RDI-based techniques. 
\item \emph{Unexpectedly low albedos}: the undetected disks may have dust compositions with albedo lower than what we expect from basic assumptions (e.g. water ice, Mie theory...), and than what we observed for the bright disks well-characterized so far.
\end{enumerate}
We should also mention the possibility of chance non-detection, due to azimuthal sensitivity variations (telescope spider) or azimuthal coverage of the instruments (HST STIS wedge), e.g. the detection of the edge-on disk around HD 377 with NICMOS \citep{Choquet2016}, but non-detection with STIS due to chance alignment with the wedge and with the telescope spider \citep{Krist2012}. Such unlucky configurations cannot explain the statistical trend seen in the different surveys with both ground-based telescopes and HST though.

The new generation of high-contrast imaging instruments on ground-based telescopes with extreme adaptive optics systems now offers smaller IWAs, opening the detection space toward more compact disks. Yet, only a handful of \modif{new} debris disks have been detected with these instruments \modif{so far}. These detections have been made possible by  improved sensitivity limits at large separation compared to the first generation of imagers, rather than by smaller IWAs \citep[e.g. HD 131835, HD 206893][]{Hung2015,Milli2017a}. Similarly, by pushing the sensitivity limits on HST-NICMOS data with modern post-processing techniques, we have discovered 11 debris disks, 10 of which with very low surface brightnesses from a few $100~\mu$Jy/arcsec$^2$ \citep{Soummer2014,Choquet2017} down to a few $10~\mu$Jy/arcsec$^2$ including these two around \firstD and \secD \citep[][, this work]{Choquet2016}. These combined results seem to indicate that there is an underlying population of debris disks much fainter than the population of bright debris disks discovered so far, indicative of low scattering albedos. A rigorous statistical analysis estimating the completeness of previous surveys to debris disks as function of their morphology is required to assess the properties of debris disks as a whole population. Such a study is out of the scope of the present paper though, and we only discuss below comparisons with previously detected systems.

The disks observed around \firstD and \secD share several common properties. They have very similar fractional infrared luminosities ($\sim6\times 10^{-4}$) and are seen with the same inclination of $\sim60$\degr, which makes their scattering properties directly comparable. 
In particular, these two disks appear very faint in scattered-light, having peak surface brightnesses of a few tens of $\mu$Jy/arcsec$^2$ only, and have  low scattering albedo values (10--15\%). A few other systems with similar inclinations also have comparably low scattering albedo values: 
HD92945,  HD207129, HD202628, and Fomalhaut  \citep[][respectively]{Golimowski2011,Krist2010,Krist2012,Kalas2005a}. 
Table~\ref{tab:photo} reports properties of debris disks with inclinations in the 50--70\degr{} range with published scattering albedo values estimated with eq.~\ref{eq:albedo} or with published scattering efficiencies. There are two populations of debris disks with distinct dust compositions, leading to low albedo values around 5--15\% for one population, and higher albedos around 50--70\% for the other. This distinction seems independent from the age of the system or the mass of the host star. Although the sample is incomplete, 
these populations with distinct albedos seem to be present in other inclinations ranges, e.g. HD 107146 and HD 181327 in the 20--30\degr{} range with scattering albedos of 15\% and 65\% respectively \citep{Schneider2014,Chen2014}, or HD 139664 and HD 61005 in the 80--90\degr{} bin with respective scattering albedos of 9\% and 64\% \citep{Schneider2014,Chen2014}.

 Interestingly, the majority of comets in the solar system display low albedos of 4--5\% in the visible \citep{Kolokolova2004} that can be well reproduced by mixtures of both submicron aggregates and compact particles \citep{Kolokolova2010}. On the other hand, infrared observations of the coma of the pristine Oort Cloud comet C/2012 K1 (Pan-STARRS) are well modeled by compact,  carbon-dominated grains, while it also displays a low albedo of $14\pm0.01$\% at scattering angle of $\sim35$\degr at infrared wavelengths (8--31~$\mu$m), comparable with the values that we measure for \firstD and \secD in the near-infrared \citep{Woodward2015}. This scattering angle, set by the position of the comet with respect to the Sun and the SOFIA telescope at the moment of the observations, is at the limit of the minimum scattering angles probed in these two disks. These examples indicate that low albedo dust  is common in the solar solar system, and can be explained both by solid grains, or by submicron aggregates, or by mixtures of both. Identifying the reason for low albedo values in debris disks systems cannot be achieved without combining resolved observations in several wavelengths regimes.
Different albedos values can indeed be explained not only by dust porosity but also by  different chemical compositions or by different dust size distributions. 
In the former case, this may indicate different initial conditions or formation mechanisms in the primordial disk that would generate planetesimals with different compositions. In the latter case, different dust size distributions would indicate different dynamical mechanisms at work after transition from the protoplanetary stage to the evolved stage of debris disks.
 A better characterization of these systems would be needed to discriminate one scenario from the other, for instance by constraining their dust size distribution with multi-band imaging.

\begin{deluxetable}{lcrrrrr}
\tablecaption{Scattering albedos of $\sim60\degr$-inclination debris disks  \label{tab:photo}}
\tablehead{
 \colhead{System}	& \colhead{Spectral}	&\colhead{Age}	& \colhead{Inc.} 	& \colhead{Scattering}	& \colhead{$\lambda$} & \colhead{Ref.}\\
  \colhead{}		& \colhead{Type}	&\colhead{(Myr)}& \colhead{(\degr)}	& \colhead{Albedo}		& \colhead{($\mu$m)} & \colhead{}
 }
\startdata
HD 202628	&G5	&2300	&64			&0.05	&0.5		&1\\
Fomalhaut 	&A4	&440		&66			&0.05	&0.8		&2\\
HD 207129	&G0	&2100	&60			&0.06	&0.6		&3\\ 
HD 92945		&K1	&294		&62			&0.09	&0.5		&4,5\\
			&	&		&			&0.10	&0.6		&6\\
\firstD		&F8	&32		& 58			&0.10	&1.1		&8	\\
\secD		&F0	&830		& 59			&0.13	&1.1		&8\\
HD 202917	&G7	&45		&69			&0.50	&0.5		&7\\
HD 15745		&F2	&23		&67			&0.63	&0.5		&4,5\\	
\enddata
\tablerefs{1: \citealt{Krist2012}; 2: \citealt{Kalas2005a}; 3: \citealt{Krist2010};  4: \citealt{Schneider2014}; 5: \citealt{Chen2014};6:\citealt{Golimowski2011}; 7: \citealt{Schneider2016}; 8: This work.}
\end{deluxetable}

\section{Conclusion}

\begin{figure}
\center
\includegraphics[width=8.5cm]{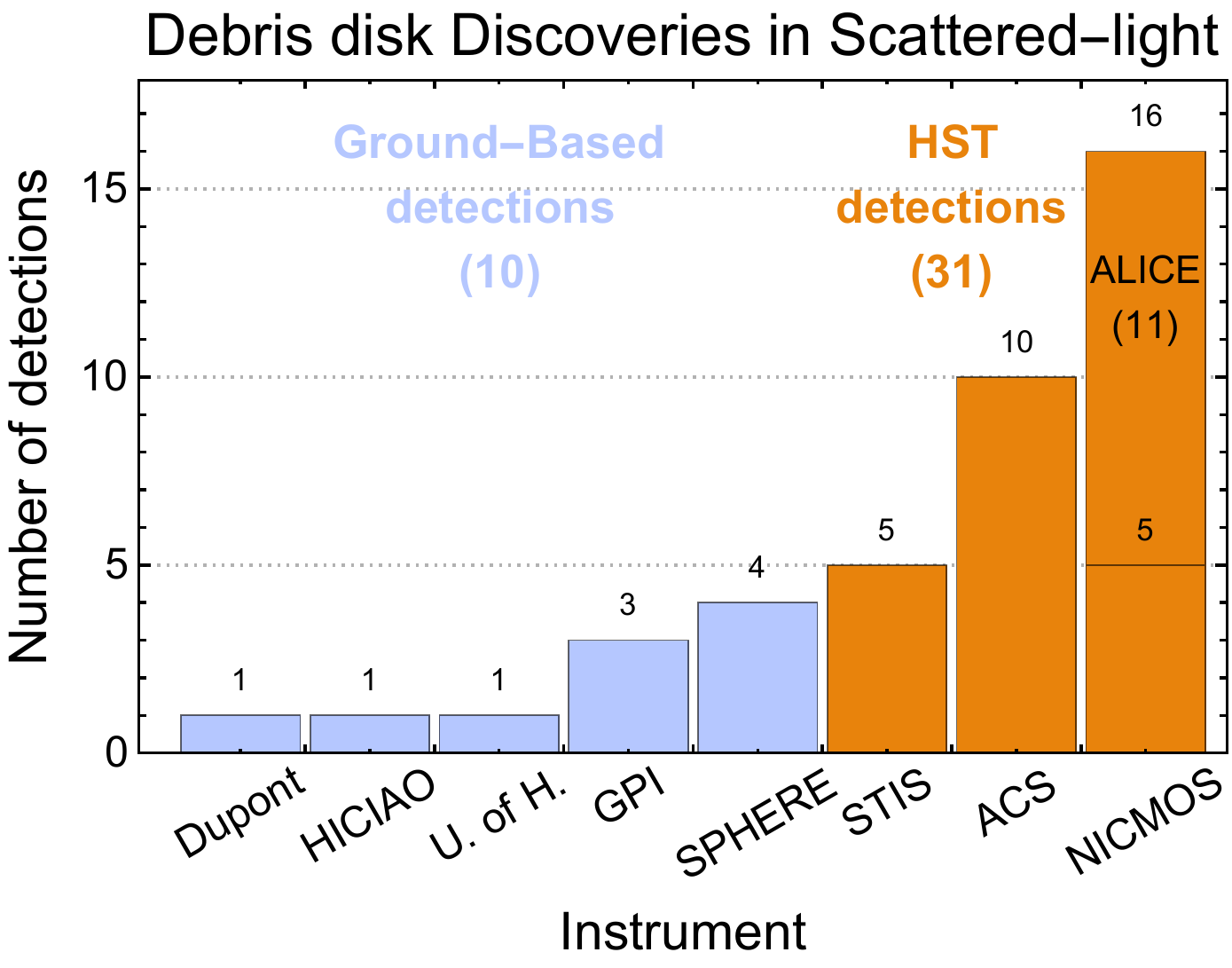}
\caption{Number of debris disks imaged for the first time in scattered-light per instrument. The ground-based instruments are highlighted in blue (total of 10 discoveries to date) and the HST instruments are highlighted in orange (31 discoveries to date). The NICMOS detections are split between pre-ALICE disks (5 detections) and ALICE disks (11 detections).\label{fig:instrudiscovery}}
\tablerefs{Dupont 2.5~m telescope: \citealt{smith1984}; 
Subaru-HICIAO: \citealt{Thalmann2013}; 
University of Hawai 2.2~m telescope: \citealt{Kalas2004}; 
Gemini-GPI: \citealt{Hung2015,Currie2015,Kalas2015}; 
VLT-SPHERE: \citealt{Kasper2015,Wahhaj2016,Matthews2017,Milli2017a}; 
HST-STIS: \citealt{Krist2012,Padgett2016}; 
HST-ACS: \citealt{Ardila2004,Kalas2005a,Kalas2006,Kalas2007a,Kalas2007b,Krist2010,Golimowski2011}, and two unpublished detections presented at conferences: HD 10647 (Stapelfeldt et al. 2007), HD 202917 (Krist et al. 2007); 
HST-NICMOS: \citealt{Weinberger1999,Augereau1999b,Schneider1999,Schneider2005,Schneider2006,Hines2007,Soummer2014,Choquet2016,Choquet2017}, this work.}
\end{figure}

To conclude, we have detected two debris disks in scattered-light, around \firstD and \secD. The former disk has previously been imaged in thermal emission with Herschel but never in scattered-light, and the latter disk has never been imaged before. These disks were found in archival HST-NICMOS data in the near-infrared, from our re-analysis using modern PSF subtraction techniques as part of the ALICE project. These two detections bring the number of debris disks discovered in scattered-light by this program to 11, and make NICMOS  the instrument with the largest number of debris disk imaged in this regime to date (see Fig.~\ref{fig:instrudiscovery}).

We carefully characterized the morphology of these disks with forward modeling techniques in order to calibrate the post-processing artifacts. The disk around \firstD has a well defined ring shape with sharp edges at a radius of 114~au. The slope of the outer edge is consistent with evolution models, and the sharp inner edge is likely sculpted by an unseen planet through secular resonances. Planets of a few Earth masses with small eccentricities can cause such ring-shapes in debris disk systems \citep{Lee2016}. The disk around \secD has a radius of 95~au. Both disks are inclined by $\sim60$\degr{} from face-on, and have very low albedo values of 10\% and 13\% respectively, which exclude compositions of pure water ice. 

They are reminiscent of several other disks previously detected with HST showing similarly low albedo values, around Fomalhaut, HD 202628, HD 207129, and HD 92945. These disks may have similar dust compositions, differing from the many brighter disks that have been imaged thus far in scattered light. Interestingly, comets in the solar systems also display comparably low albedo values in the visible. Porous grains, chemical composition,  as well as different dust size distribution may explain the differences in albedo  in the observed populations of disks. A better characterization of these systems with images in complementary bandpasses would help understand the different properties between these systems.

\acknowledgments
 © 2017. All rights reserved.

E.C. acknowledges support from NASA through Hubble Fellowship grant HF2-51355 awarded by STScI, which is operated by AURA, Inc. for NASA under contract NAS5-26555, for research carried out at the Jet Propulsion Laboratory, California Institute of Technology. 
J.-C.A. acknowledges support from the ``Programme National de Plan\'etologie'' (PNP) of CNRS/INSU co-funded by the CNES. 
This work is based on data reprocessed as part of the ALICE program, which was supported by NASA through grants HST-AR-12652 (PI: R. Soummer), HST-GO-11136 (PI: D. Golimowski), HST-GO-13855 (PI: E. Choquet), HST-GO-13331 (PI: L. Pueyo), and STScI Director's Discretionary Research funds. Part of these data were calibrated as part of the LAPLACE program, which was supported by NASA through grants HST-AR-11279 (PI: G. Schneider).
This research has made use of NASA's Astrophysics Data System (ADS),  the Washington Double Star Catalog, maintained at the U.S. Naval Observatory, the Keck Observatory Archive (KOA), operated by the W. M. Keck Observatory and the NASA Exoplanet Science Institute (NExScI) under contract with NASA, the SIMBAD database, operated at CDS, Strasbourg, France, and of data from the European Space Agency (ESA) mission {\it Gaia}.
This research has used archival data from HST programs HST-GO-10527 (PI: D. Hines) and HST-GO-11157 (PI: J. Rhee), and from Keck programs. We thank the anonymous referee for her or his comments which made the paper much clearer.

 \facilities{HST(NICMOS)}.

\bibliographystyle{aasjournal} 
\bibliography{/Users/echoquet/Documents/Biblio/biblio-disk-planet.bib}

\end{document}